\documentclass[review,11pt]{elsarticle}

\usepackage{lineno,hyperref}
\usepackage{amsmath}
\usepackage{mhchem}
\usepackage{prettyref}
\usepackage{subfigure}
\usepackage{float}
\usepackage{booktabs}
\usepackage[margin=3.2cm]{geometry}

\hypersetup{hyperindex=false, colorlinks=true, linkcolor=blue, citecolor=blue, urlcolor=blue}
\newrefformat{fig}{Fig.~\ref{#1}}
\newrefformat{tbl}{Table~\ref{#1}}
\newrefformat{sec}{Section~\ref{#1}}
\newrefformat{eq}{Eq.~\ref{#1}}
\newrefformat{alg}{Algorithm~\ref{#1}}

\makeatletter
\def\ps@pprintTitle{%
   \let\@oddhead\@empty
   \let\@evenhead\@empty
   \def\@oddfoot{\reset@font\hfil\thepage\hfil}
   \let\@evenfoot\@oddfoot
}
\makeatother

\usepackage{numcompress}

\begin{document}

\begin{frontmatter}

\title{Historical shear deformation of rock fractures derived from digital outcrop models and its implications on the development of fracture systems\tnoteref{t1}}

\tnotetext[t1]{Code available from: https://github.com/EricAlex/structrock.}

\author[zju]{Xin Wang}

\author[zju]{Yi Qin}

\author[zju]{Zhaohui Yin}

\author[zju]{Lejun Zou}

\author[zju]{Xiaohua Shen}

\address[zju]{School of Earth Sciences, Zhejiang University, Hangzhou, China}




\begin{abstract}

The initiation and development of fractures in rocks is the key part of many problems from academic to industrial, such as faulting, folding, rock mass engineering, reservoir characterization, etc. Conventional ways of evaluating the fracture historical deformations depend on the geologists' visual interpretation of indicating structures such as fault striations, fault steps, plumose structures, etc. on the fracture surface produced by previous deformations, and hence suffer from problems like subjectivity and the absence of obvious indicating structures.  In this study, we propose a quantitative method to derive historical shear deformations of rock fractures from digital outcrop models (DOMs) based on the analysis of effects of fault striations and fault steps on the shear strength parameter of the fracture surface. A theoretical model that combines effects of fault striations, fault steps and isotropic base shear strength is fitted to the shear strength parameter. The amount of fault striations and fault steps and their occurrences are estimated, and the historical shear deformations can be inferred. The validity and the effectiveness of the proposed method was proved by testing it on a constructed fracture surface with idealized striations and a fracture surface with clear fault steps. The application of this method on an example outcrop shows an intuitive idea of how the rock mass was deformed and that the distribution, occurrence and mode of new fractures are strictly controlled by preexisting fractures, and hence emphasizes the importance of preexisting fractures in modeling the development of fracture systems.

\end{abstract}


\end{frontmatter}


\section{Introduction}

Rock fractures in the field usually have complex mechanical origin and deformation history. The complex relationships between fracture deformation, fracture network, faulting, folding, rock mass properties, reservoir production, geological \ce{CO2} storage, and etc. make the fracture deformation data the key part of many problems from academic to industrial. However, to our knowledge, the fracture mechanical origin and historical deformations were mainly inferred from the geologists' visual interpretation of the indicating structures -- if they were very well shown -- such as fault striations, fault steps, plumose structures, offsets, etc. Beside this method's drawbacks of requiring large amount of labor and prone to subjectivity when the indicating structures are not so obvious for visual interpretation, quite commonly, the fracture doesn't show any obvious indicating structures for visual interpretation at all. Among those types of indicating structures, plumose structures are produced by tensile deformation of rock fractures; offsets are found on cross-sections of fault planes; while fault striations and fault steps are indicating structures on the fault plane produced by brittle shear deformations. In this paper, we are interested in fault striations, fault steps and shear deformation of rock fractures, so in the later text, we use ``indicating structures'' to refer in particular to fault striations and fault steps for simplicity.

The Digital Outcrop Model (DOM) \citep{Bellian05}, which is usually acquired using techniques such as laser scanning, structure from motion or stereo vision, describes the outcrop surface in the form of a point cloud, a large number of 3D $(X, Y, Z)$ points, and thereby contains morphological features indicating historical deformations of rock fractures. The 3D geometries of those indicating structures may vary wildly depending on the magnitude of the shear deformation, the properties of the rock, and so on. An important fact is that a certain type of indicating structure can have similar effects on the shear strength of a fracture. For instance, the existence of fault steps would make it harder for the fracture to shear in the direction that faces the steps. Thus it's possible to estimate the existence and the amount of indicating structures, hence the historical shear deformations, by analyzing the effects of these indicating morphological features on the shear strength of rock fractures. In doing so, the contribution of surface morphology on the shear strength of rock fractures should firstly be modeled. Efforts have been made in the literature to estimate the shear strength of rock fractures while trying to study the contribution of surface morphology. Among all the models (e.g., Ladanyi's empirical model \citep{Ladanyi1969}, Barton's empirical model \citep{Barton1977}, Amadei-Saeb's analytical model \citep{Amadei1998,Saeb1992} and Plesha's theoretical model \citep{Plesha1987}), Barton's morphological parameter known as the joint roughness coefficient (JRC) is the one that is mainly used in practice \citep{ISRM1978}. As shearing strictly depends on three-dimensional contact area location and distribution \citep{Gentier2000}, \citet{Grasselli2002} proposed a method for the quantitative three-dimensional description of a rough fracture surface, and based on this description, \citet{Grasselli2003} proposed a constitutive criterion to model the shear strength of fractures and also to estimate the JRC value. This constitutive model was reported to be able to describe experimental shear tests conducted in the laboratory \citep{Grasselli2003}. Based on this three-dimensional description, which describes the contribution of surface morphology on the shear strength of rock fractures, the effects of indicating structures can be analyzed, since indicating structures are part of the surface morphology.

On the other hand, each individual fracture surface should be extracted from outcrop point clouds, and during the last 10 years, many methods (e.g., \citep{Slob05,Roncella05,Voyat06,Olariu08,Lato09,Ferrero09,Gigli11,Garcia11,Riquelme14,Gomes16,Wang2017}) have been developed because of the wide range of applications of the fracture data. Among those methods, the region-growing approach proposed by \citet{Wang2017} is specially designed for the automatic extraction of the full extent of each individual fracture surface from the outcrop point cloud, unlike methods that do not extract the full extent of each individual fracture \citep{Slob05,Olariu08,Lato09,Gomes16} or methods that need human supervision \citep{Ferrero09,Gigli11,Garcia11,Riquelme14}. And using the region-growing method, \citet{Wang2017} reported that automatically extracted fracture data are of the same or even better quality as the manually surveyed data.

Based on researches described above, this paper present a quantitative method to derive historical shear deformations of rock fractures from DOMs. A constructed fracture surface with idealized striations and a fracture surface with clear fault steps were used to test this method. The results are consistent with the occurrence of idealized striations and visual interpretations of fault steps and therefore prove the validity of this method. The application of this quantitative method on an example outcrop shows indications of how the rock mass was deformed. The strict control of preexisting fractures on new fractures emphasizes the importance of preexisting fractures in the modeling of the development of fracture systems.

\section{Methodology}

\subsection{The extraction of individual fracture surfaces from outcrop point clouds}

\begin{figure}[h]
  \centering
  \subfigure[]
  {\includegraphics[width=0.8\textwidth]{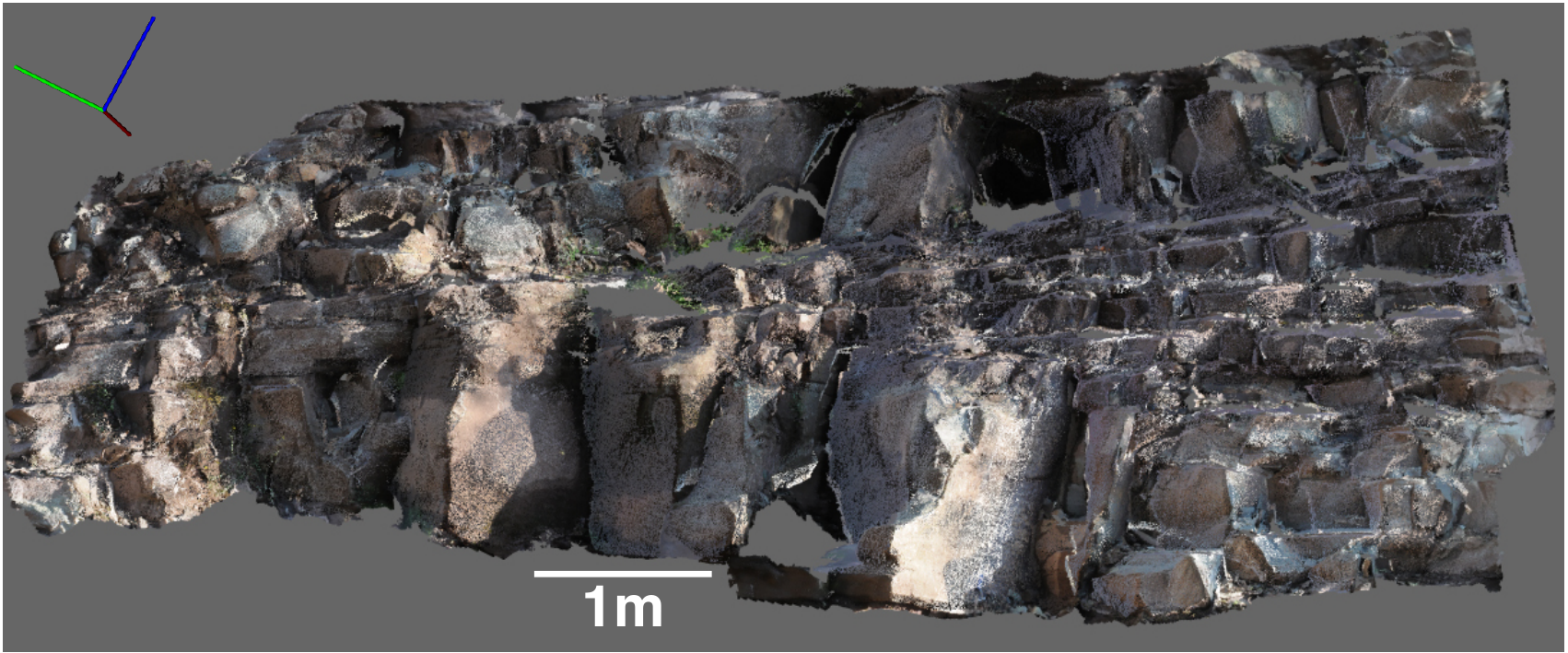}}
  \subfigure[]
  {\includegraphics[width=0.8\textwidth]{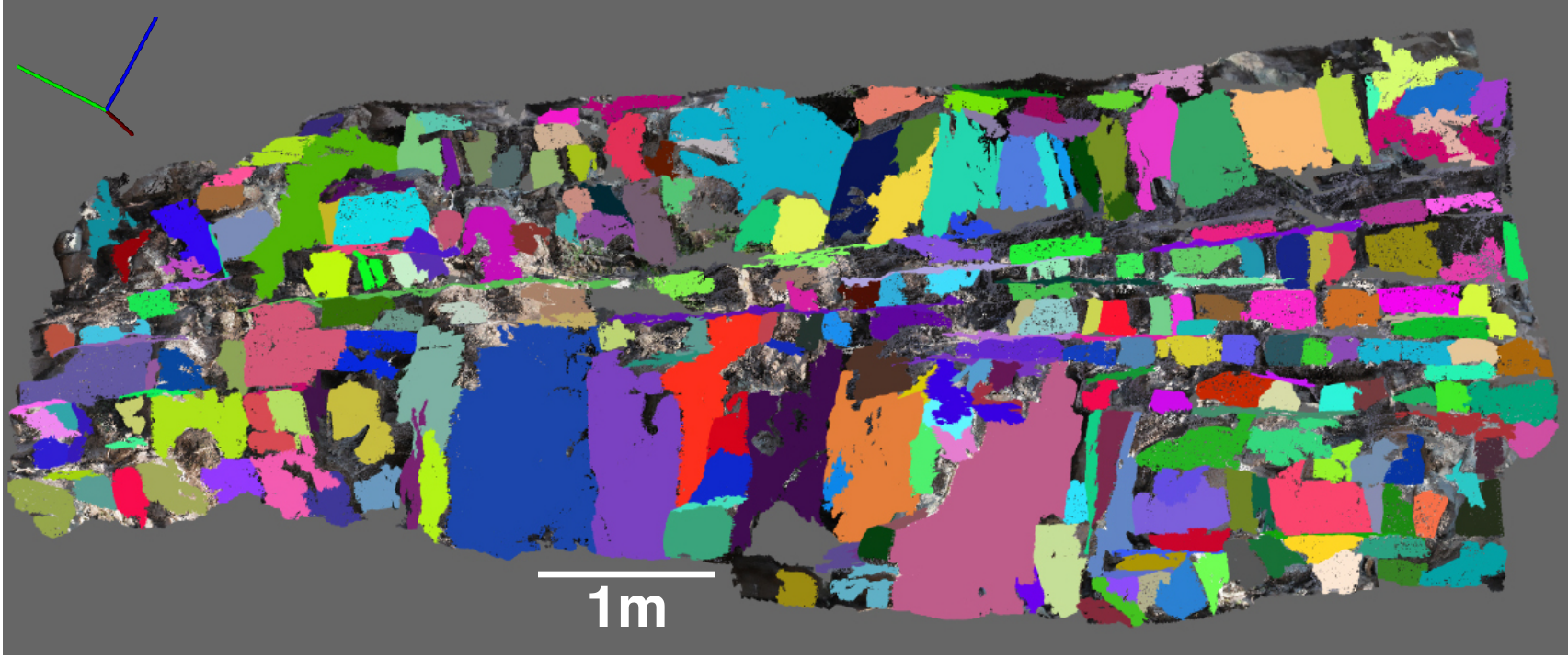}}
  \caption{(a) The point cloud scanned from an example outcrop (see \prettyref{sec:applications_and_discussion} for details) and (b) the result of individual fracture surfaces extraction from the outcrop point cloud. Different fracture surfaces are shown by different colors. The reference coordinate system is in the upper-left corner: the green axis points north, the red axis points east and the blue axis points vertically up.}
  \label{fig:point_cloud_fracture_faces}
\end{figure}

Using laser scanning instruments, an outcrop (see \prettyref{sec:applications_and_discussion} for details) is scanned as point clouds (\prettyref{fig:point_cloud_fracture_faces}a). Rock fracture surfaces need to be extracted from outcrop point clouds before any analysis can be conducted. And in order to account for all features on the fracture surface, the full extent of individual fracture surfaces should be extracted. The region-growing approach \citep{Wang2017} was specially designed for this purpose and was proven to be able to get high quality results, as shown in \prettyref{fig:point_cloud_fracture_faces}b, in which different fracture surfaces are shown by different colors.

\subsection{The contribution of surface morphology on the shear strength of rock fractures}
\label{sec:quantitative_three-dimensional_description}

The indicating structures on fracture surfaces are in fact the results of historical shear deformations. Although the 3D geometries of those indicating structures may vary wildly depending on the magnitude of the shear deformation, the properties of the rock, and so on, a certain type of indicating structure can have similar effects on the shear strength of a fracture. Thus it's possible and logically reliable to estimate the existence and the amount of indicating structures by deciphering the shear strength of fractures. Note that it's not necessary to use the real shear strength of fractures to estimate the indicating structures, and the contribution of fracture surface morphology on the shear strength is sufficient as the indicating structures are themselves morphological structures.

\begin{figure}[h]
  \centering
  \subfigure[]
  {\includegraphics[width=0.4\textwidth]{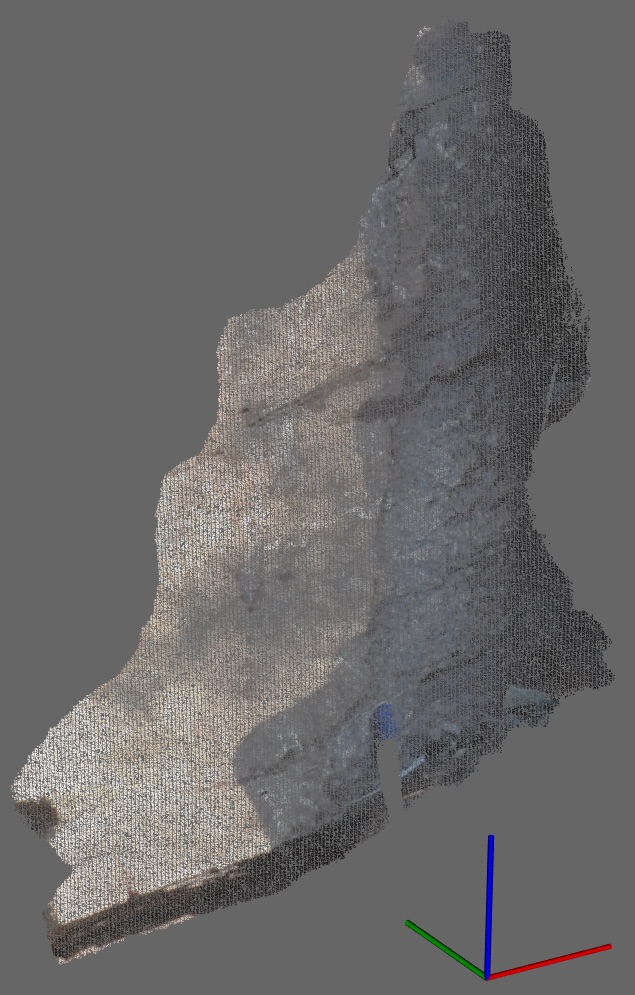}}
  \subfigure[]
  {\includegraphics[width=0.402\textwidth]{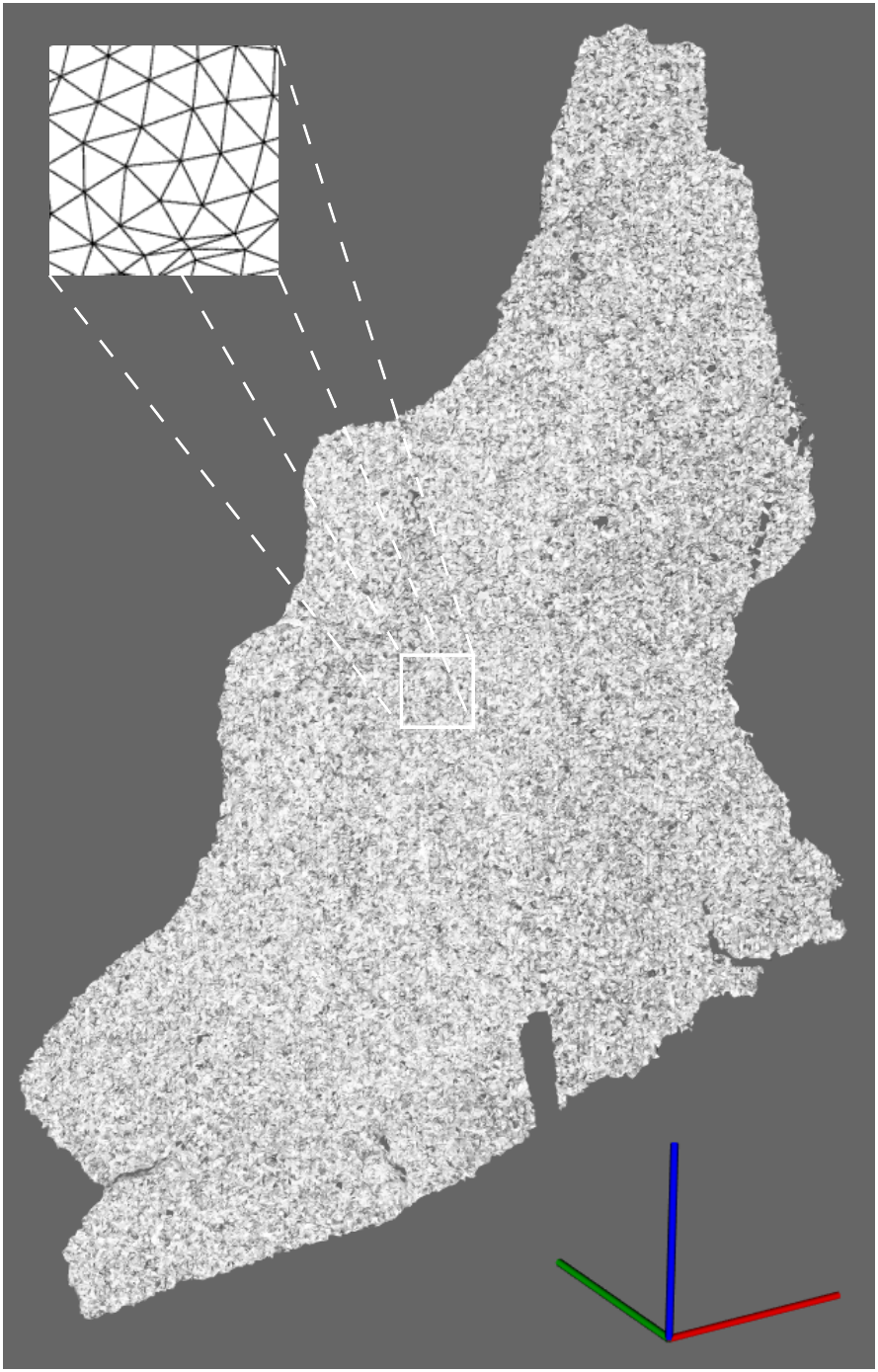}}
  \caption{(a) The point cloud of a fracture surface and (b) the reconstructed fracture surface using a triangulation algorithm proposed by \citet{Marton09ICRA}. The reference coordinate system is in the lower-right corner: the green axis points north, the red axis points east and the blue axis points vertically up.}
  \label{fig:before_n_after_triangulation}
\end{figure}

The triangulated surfaces were reconstructed using a triangulation algorithm proposed by \citet{Marton09ICRA} on fracture point clouds in order to handle the ``fractal'' structures of fracture surfaces (\prettyref{fig:before_n_after_triangulation}). The parameter $\theta_{max}^*/C$ that describes the contribution of fracture surface morphology on the shear strength was proposed by \citet{Grasselli2002}. The parameter is able to capture and quantify the effect of surface anisotropy on the shear strength of a fracture \citep{Grasselli2003}.

In the calculation of $\theta_{max}^*/C$, the potential contact area $A_{\theta^*}$, the sum of all areas in contact or damaged during shearing, is very important as shearing strictly depends on three-dimensional contact area location and distribution. With specific shear direction $\mathbf{t}$, for each triangle surface, the azimuth angle $\alpha$ is the angle between the true dip vector
projection on the shear plane $\mathbf{w}$ and the shear vector $\mathbf{t}$, measured clockwise from $\mathbf{t}$, the dip angle $\theta$ is the angle between the shear plane and the triangle, the apparent dip angle $\theta^*$ describes the contribution of
each triangle inclination, where
\begin{linenomath*}
\begin{equation}
 \tan \theta^* = - \tan \theta \cos \alpha.
\end{equation}
\end{linenomath*}
The identification of the potential damaged areas only requires to determine the areas which face the shear direction and which, among them, are steep enough to be involved \citep{Grasselli2002}. In a simplified shearing mechanism \citep{Grasselli2002}, $A_{\theta^*}$ is the sum of all areas of the
surface facing the shear direction, and steeper than a threshold apparent inclination (denoted as $\theta_{cr}^*$ ). To describe the relationship between the potential contact area $A_{\theta^*}$ and the corresponding minimum apparent dip angle $\theta^*$, the following equation \citep{Grasselli2002} was adopted to fit the data:
\begin{linenomath*}
\begin{equation}
 A_{\theta^*} = A_0 \left(\frac{\theta_{max}^* - \theta^*}{\theta_{max}^*}\right)^C,
\end{equation}
\end{linenomath*}
where $A_0$ is the maximum possible contact area in the shear direction, $\theta_{max}^*$ is the maximum apparent dip angle in the shear direction, and $C$ is a ``roughness'' parameter, calculated using a best-fit regression function, which characterizes the distribution of the apparent dip angles over the surface.

\begin{figure}[H]
  \centering
  \subfigure[]
  {\includegraphics[width=0.45\textwidth]{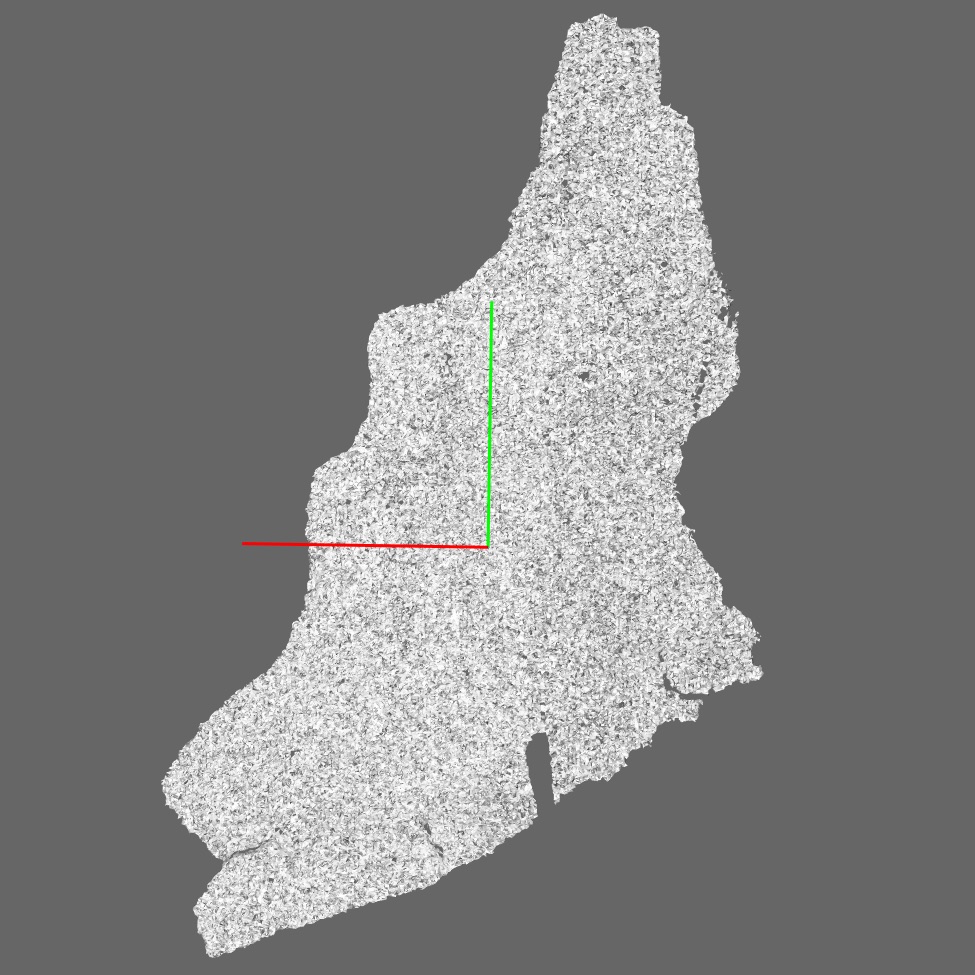}}
  \subfigure[]
  {\includegraphics[width=0.45\textwidth]{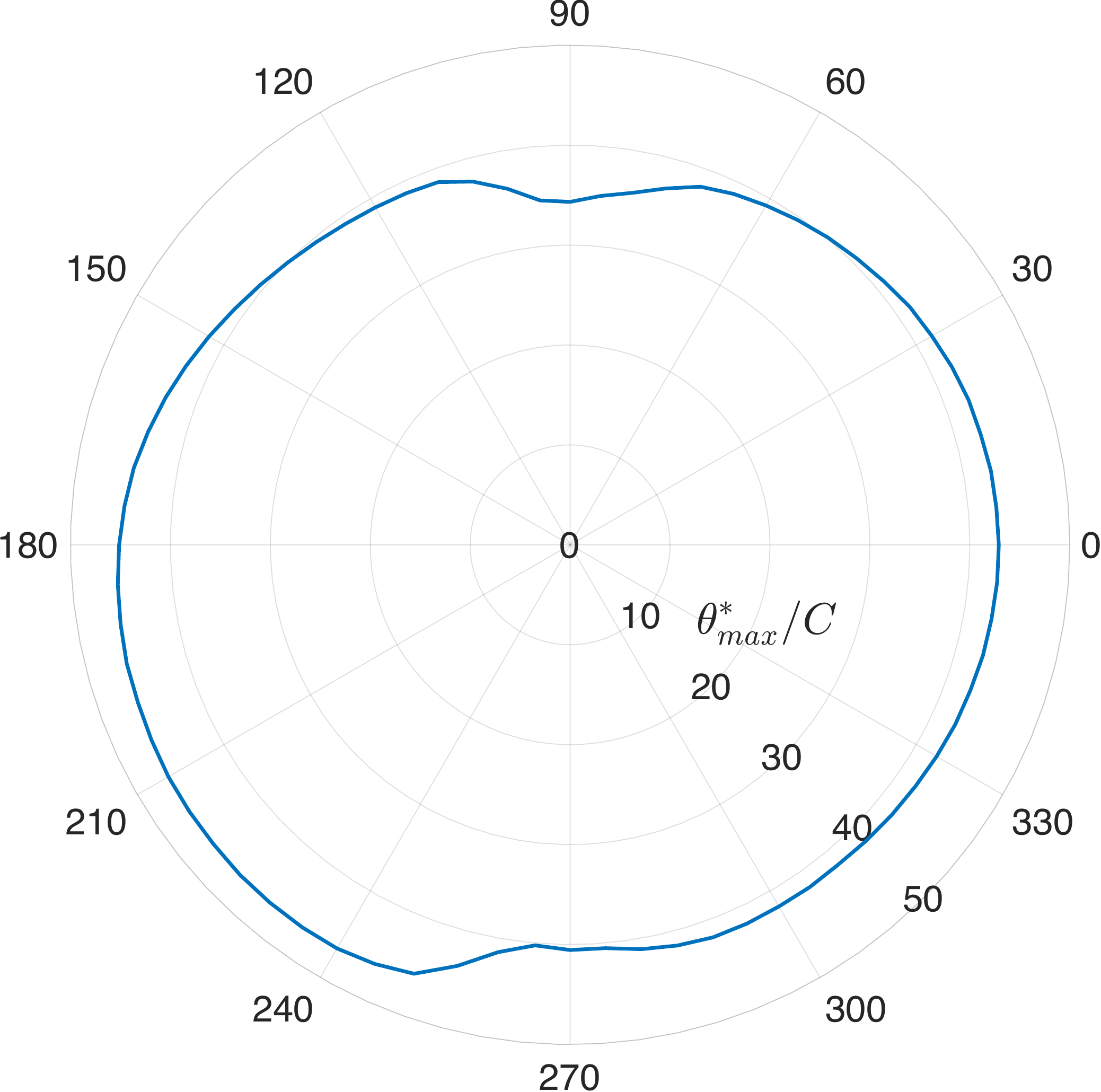}}
  \caption{(a) The reconstructed fracture surface. (b) The polar diagram of $\theta_{max}^*/C$ calculated for shear directions all around the average plane of the reconstructed fracture surface. For reference, the red line in (a) points in the shear direction of $0^\circ$ and the green line points in the shear direction of $90^\circ$. Note the anisotropic distribution of values $\theta_{max}^*/C$ for the fracture surface.}
  \label{fig:after_triangulation_n_polar_diagram}
\end{figure}

The ratio $\theta_{max}^*/C$ describes the change of angularity across the surface; low values indicate relatively few steeply inclined areas, which, therefore, corresponds to less shear strength. And the shear-strength-related parameter $\theta_{max}^*/C$ can be obtained for each shear direction on the fracture surface to capture and quantify the effect of surface anisotropy (such as the indicating structures) on the shear strength of a fracture. For instance, the values of $\theta_{max}^*/C$ were calculated for shear directions all around the average plane of the fracture shown in \prettyref{fig:before_n_after_triangulation} in steps of $5^\circ$. The polar diagram of $\theta_{max}^*/C$ shows the anisotropic behavior in shear strength of this rock fracture (\prettyref{fig:after_triangulation_n_polar_diagram}b).

\subsection{Theoretical model of effects of indicating structures and the estimation of historical shear deformations of rock fractures}

We assume that for the same rock type and similar normal load on the fracture, the amount of indicating structures is positively associated with the degree of historical shear deformations. And we know that the amount of indicating structures is positively associated with the degree of anisotropic behavior in shear strength of the rock fracture. So the problem becomes the estimation of the amount and the spatial occurrence of indicating structures from the polar diagram of $\theta_{max}^*/C$. Note that the indicating structures, i.e. ``fault striations'' and ``fault steps'', we are talking about here are not necessarily as obvious as what a geologist would have considered standard --- they are estimated as long as they are causing anisotropic behavior in shear strength of the rock fracture. Here we call them quasi fault striations and quasi fault steps.

\begin{figure}[H]
  \centering
  \subfigure[Isotropic ``base'']
  {\includegraphics[width=0.46\textwidth]{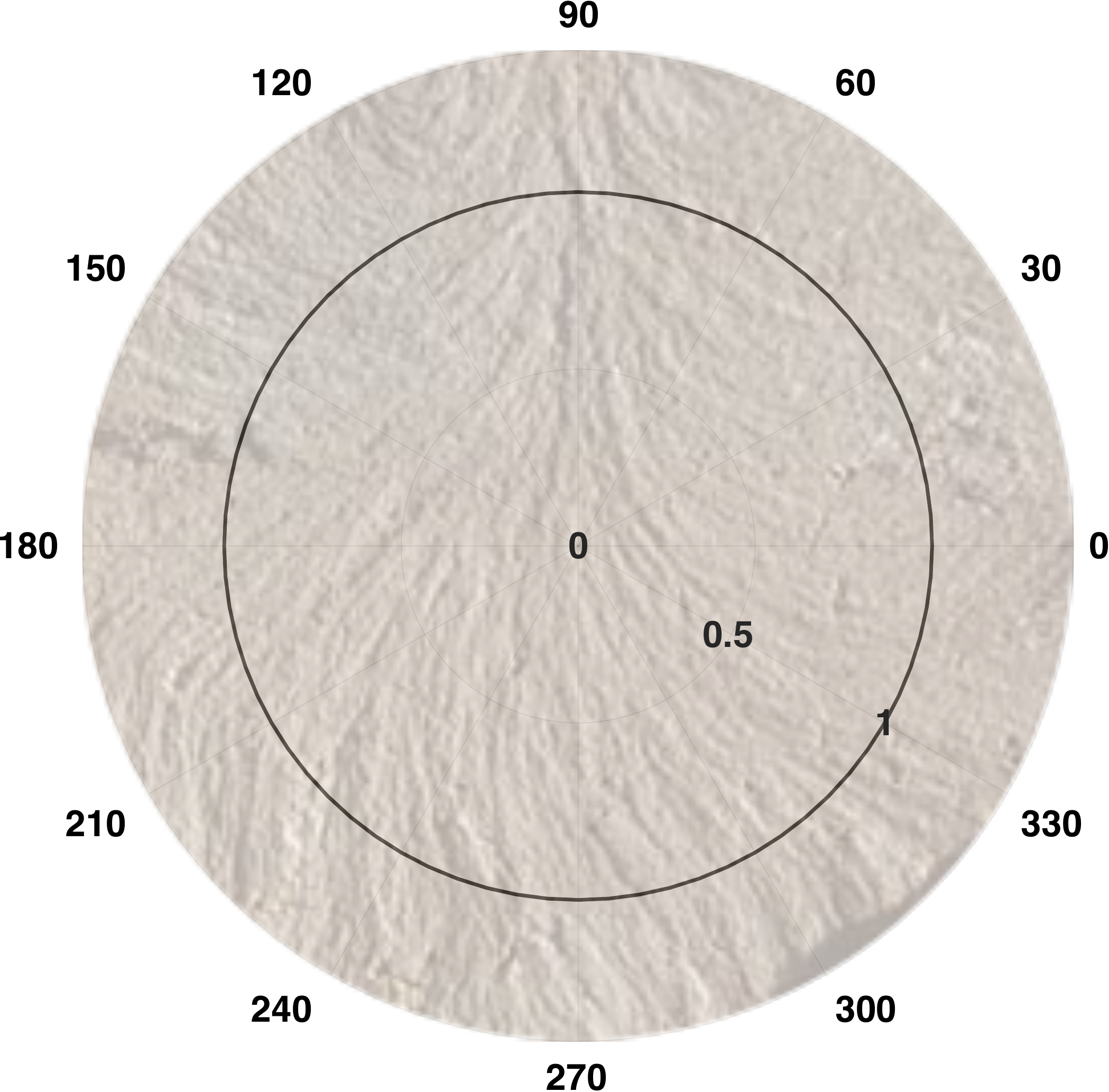}}
  \subfigure[Effects of quasi striations]
  {\includegraphics[width=0.53\textwidth]{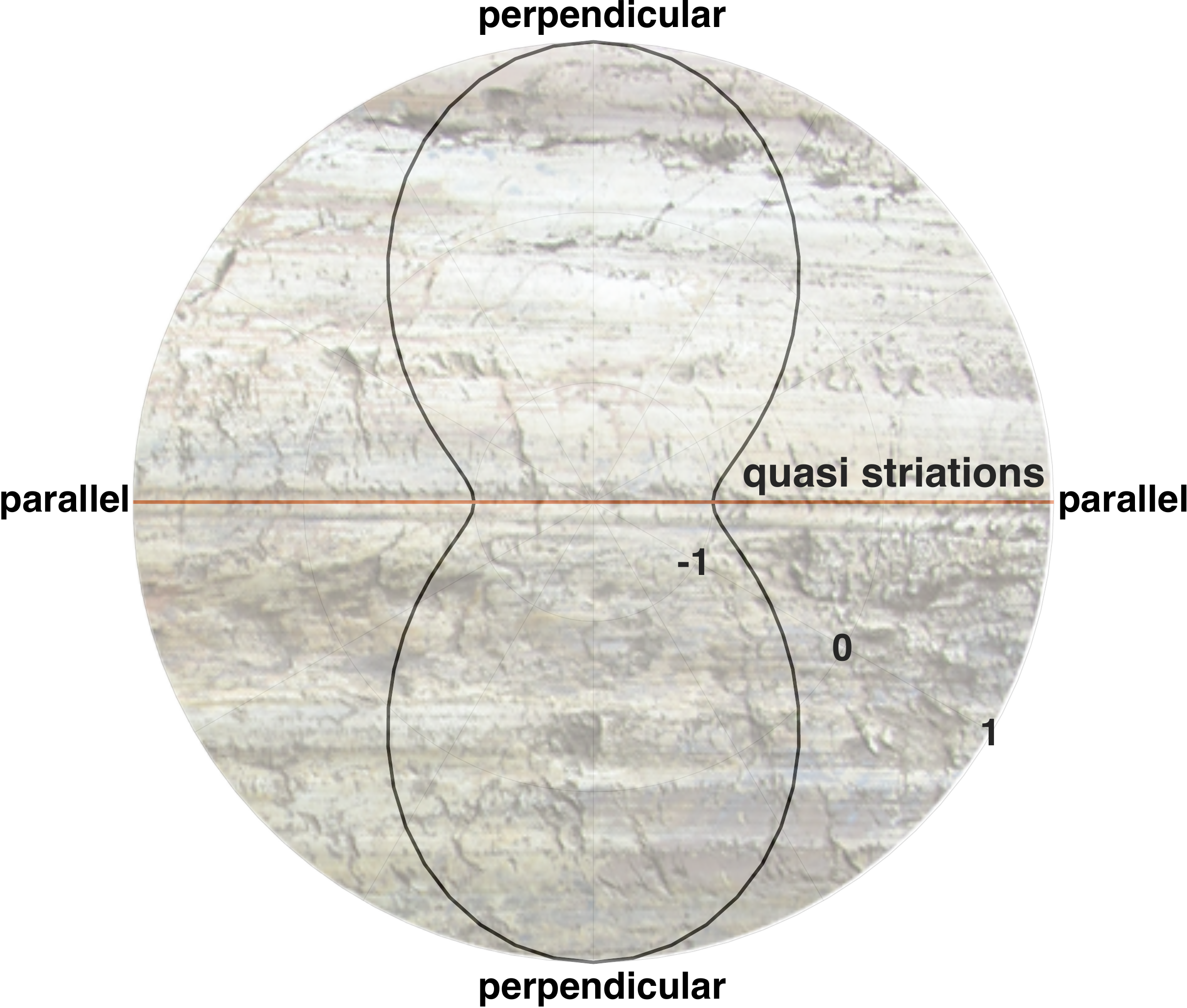}}
  \subfigure[Effects of quasi steps]
  {\includegraphics[height=0.45\textwidth]{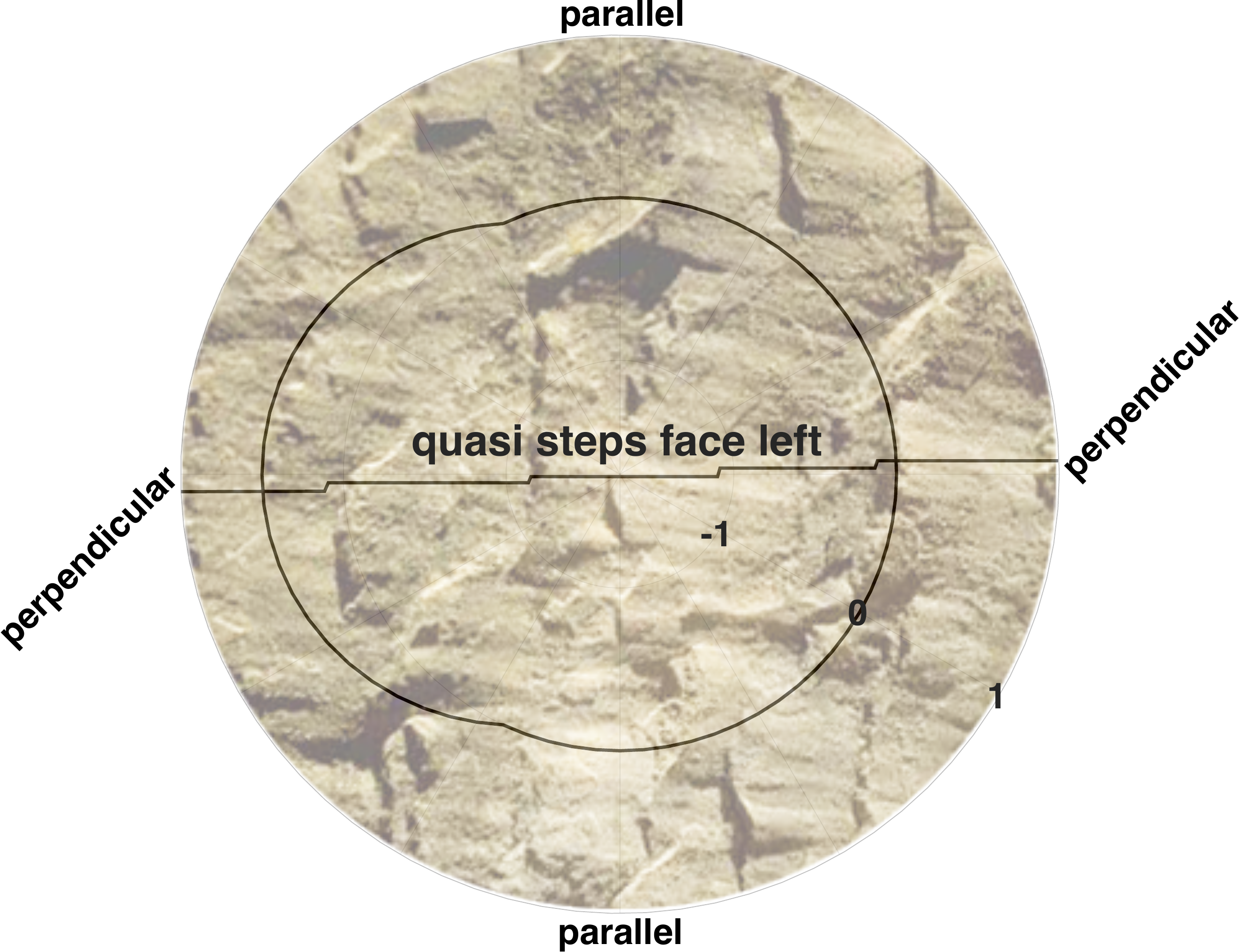}}
  \caption{The illustration of effects of indicating structures on the shear strength of rock fractures. (a) The isotropic ``base'' shear strength parameter without effects of anisotropic structures. (b) The ``sinusoidal'' effects of quasi striations. (c) The strengthening effects of quasi steps.}
  \label{fig:base_striations_steps_effects}
\end{figure}

When considering the effects of quasi fault striations and quasi fault steps on the shear strength of rock fractures, it's easy to understand that the quasi fault striations have ``sinusoidal'' effects --- the shear strength is weakened in directions parallel to the quasi striations while strengthened in directions perpendicular to the quasi striations (\prettyref{fig:base_striations_steps_effects}b), and that the shear strength is strengthened in shear directions that face the quasi steps (\prettyref{fig:base_striations_steps_effects}c). On the other hand, if the quasi striations and quasi steps, hence their anisotropic effects, are removed, we have the isotropic ``base'' shear strength parameter of rock fractures (\prettyref{fig:base_striations_steps_effects}a).

We can formulate the effects of quasi fault striations as:
\begin{linenomath*}
\begin{equation}
 f(\phi) = M_f \sin \left\{2 \left[\phi - \left(\phi_f - \frac{3\pi}{4}\right)\right]\right\},
\end{equation}
\end{linenomath*}
where $M_f > 0$ is the magnitude of the quasi striations' effects and $\phi_f$ is the direction that parallels to the quasi striations. The effects of quasi fault steps can be formulated as:
\begin{linenomath*}
\begin{eqnarray}
 g(\phi) & = & M_g \Pi\left[\frac{2(\phi - \phi_g)}{T}\right] \sin\left\{\frac{2\pi}{T}\left[\phi - \left(\phi_g - \frac{T}{4}\right)\right]\right\} \nonumber \\
 && + M_g \Pi\left\{\frac{2[\phi - (\phi_g + 2\pi)]}{T}\right\} \sin\left\{\frac{2\pi}{T}\left[\phi - \left(\phi_g + 2\pi - \frac{T}{4}\right)\right]\right\},
\end{eqnarray}
\end{linenomath*}
where $M_g >0$ is the magnitude of the quasi steps' effects, $\Pi$ is the rectangular function defined as:
\begin{linenomath*}
\begin{equation}
    \Pi(t)= 
\begin{cases}
    0, & \text{if } |t| > \frac{1}{2}\\
    \frac{1}{2},& \text{if } |t| = \frac{1}{2}\\
    1, & \text{if } |t| < \frac{1}{2}
\end{cases}
\end{equation}
\end{linenomath*}
$T$ is the period of the sine function, $\phi_g \in (\frac{T}{4}-2\pi, 2\pi-\frac{T}{4})$ is the direction the quasi steps face. Due to the fact that quasi steps are usually not strictly facing the same direction, the range of $\phi$ over which the quasi steps have effects, i.e., $T/2$, is larger than $\pi/2$. The value of $T/2$ is usually set in the range of $(\pi/2, \pi)$ depending on the occurrence of the quasi steps.

\begin{figure}[H]
  \centering
  \includegraphics[width=0.8\textwidth]{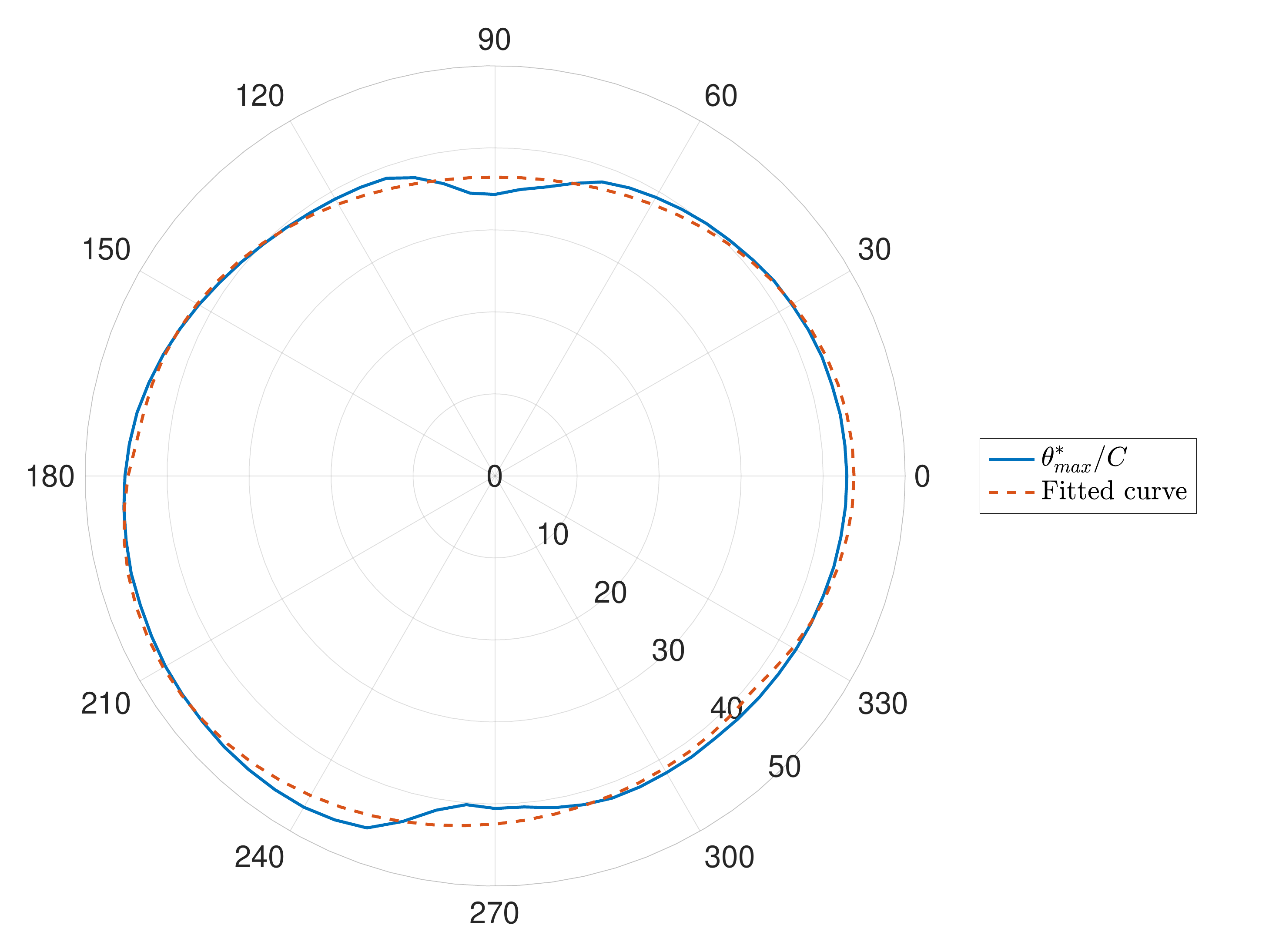}
  \caption{Fitting model $F(\phi)$ to the polar diagram of $\theta_{max}^*/C$. Estimated parameters of the fitting model: 0.0916 ($M_f/B$), $90^\circ$ ($\phi_f$), 0.1699 ($M_g/B$) and $247.17^\circ$ ($\phi_g$).}
  \label{fig:model_fitted}
\end{figure}

Thus we have a model $F(\phi)$ to describe the shear-strength-related parameter $\theta_{max}^*/C$:
\begin{linenomath*}
\begin{equation}
F(\phi) = B + f(\phi) + g(\phi),
\end{equation}
\end{linenomath*}
where $B$ is the base shear strength parameter of rock fractures. Fitting this model to the polar diagram of $\theta_{max}^*/C$ obtained in  \prettyref{sec:quantitative_three-dimensional_description}, as shown in \prettyref{fig:model_fitted}, we have the estimations of $B$, $M_f$, $\phi_f$, $M_g$ and $\phi_g$ for this fracture surface, i.e., we get the amount and the spatial occurrence of quasi striations and quasi steps on this fracture surface. For quasi striations, the amount is $M_f/B$ and the spatial occurrence is $\phi_f$, and for quasi steps, the amount is $M_g/B$ and the spatial occurrence is $\phi_g$.

\section{Method validation and performance analysis with different point densities of the point cloud}

\begin{figure}[H]
  \centering
  \subfigure[]
  {\includegraphics[width=0.49\textwidth]{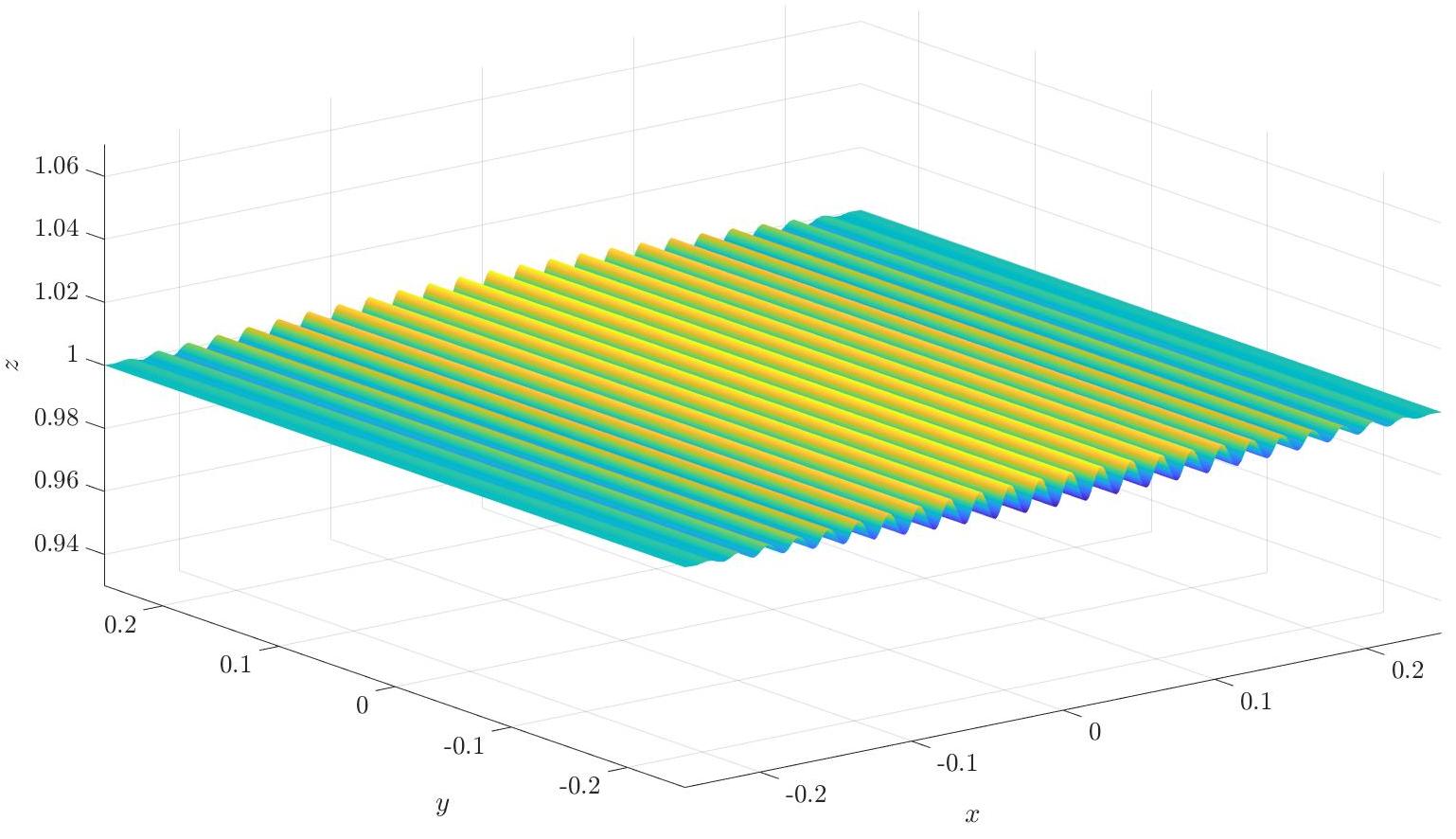}}
  \subfigure[]
  {\includegraphics[width=0.49\textwidth]{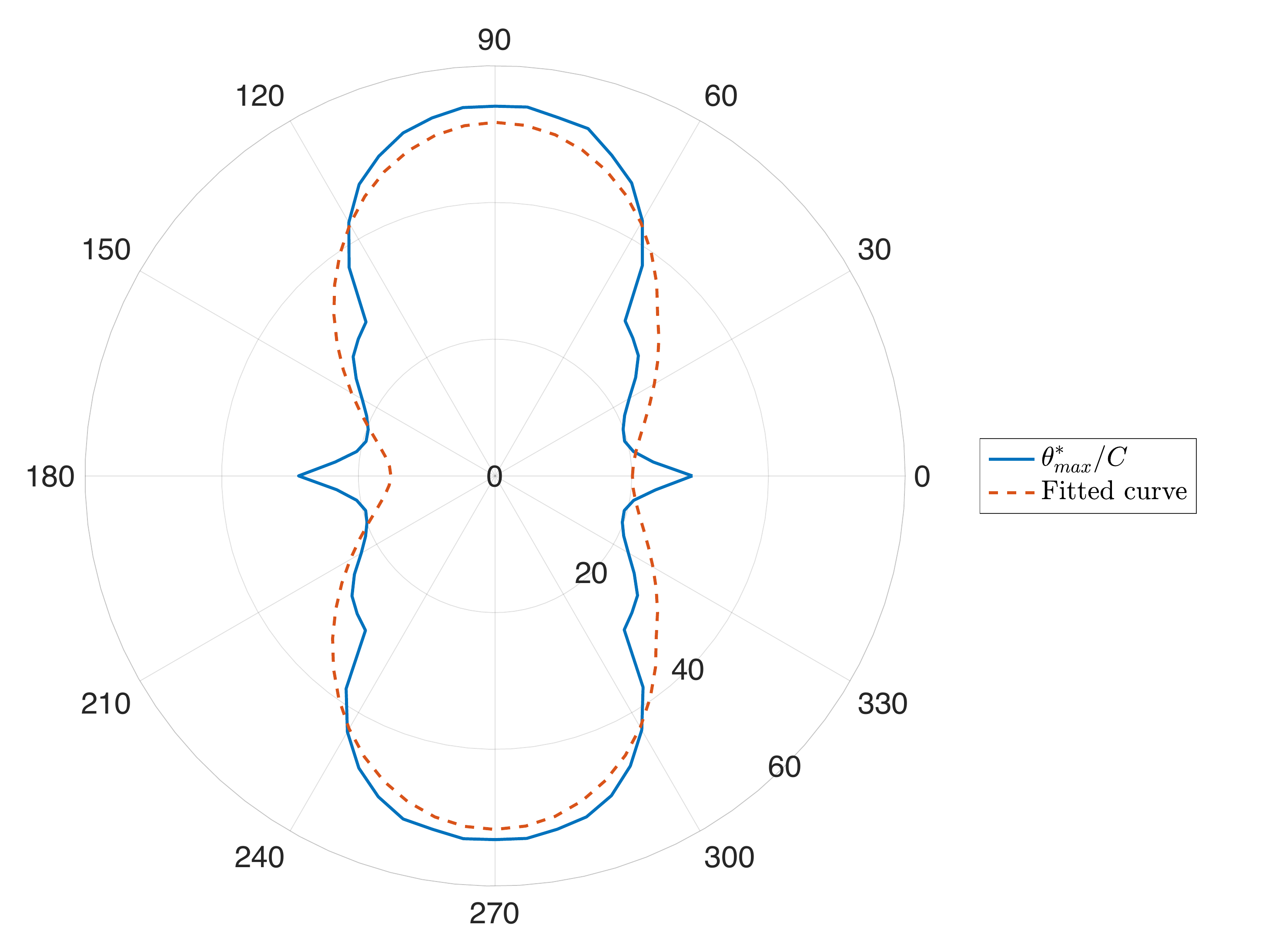}}
  \subfigure[]
  {\includegraphics[width=0.32\textwidth]{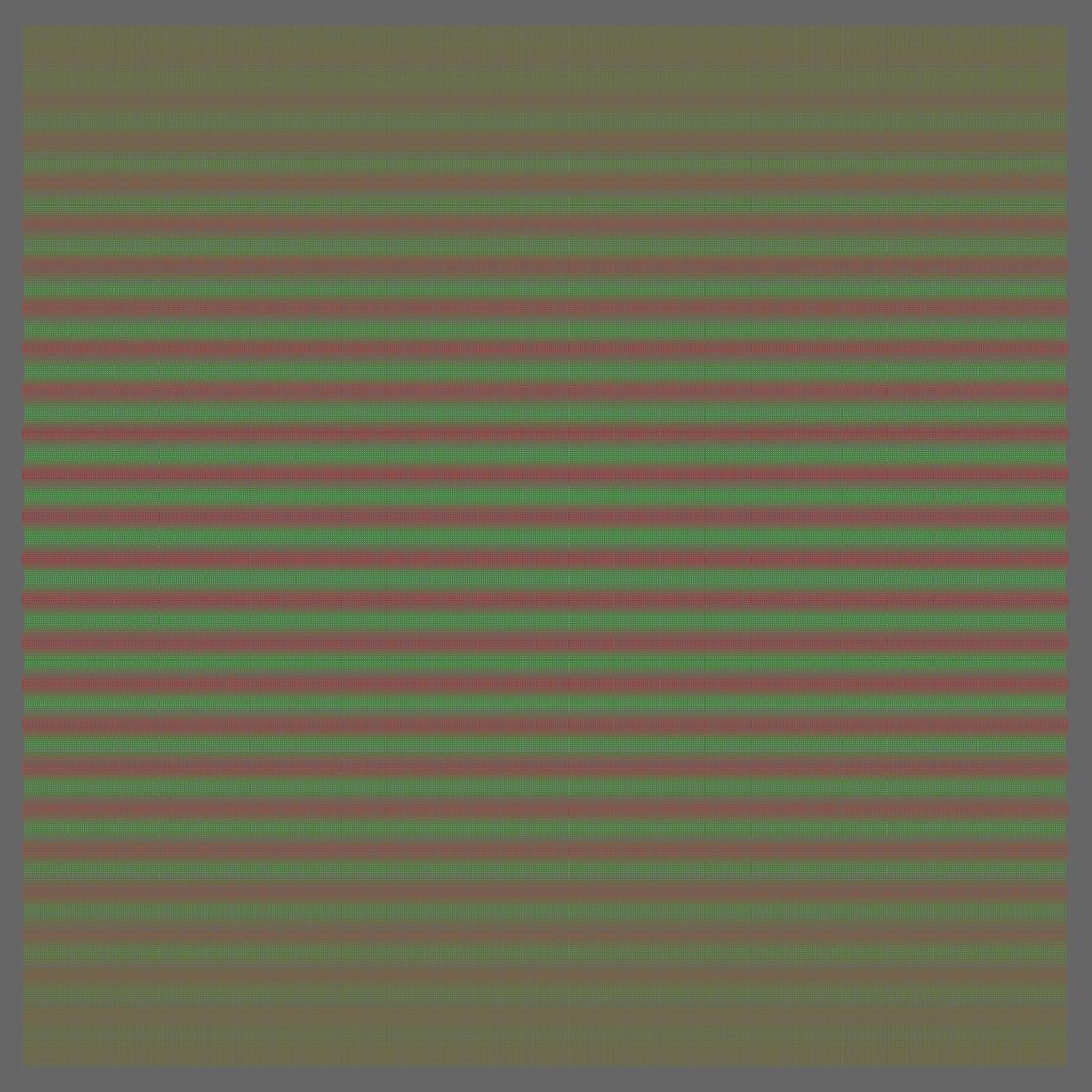}}
  \subfigure[]
  {\includegraphics[width=0.32\textwidth]{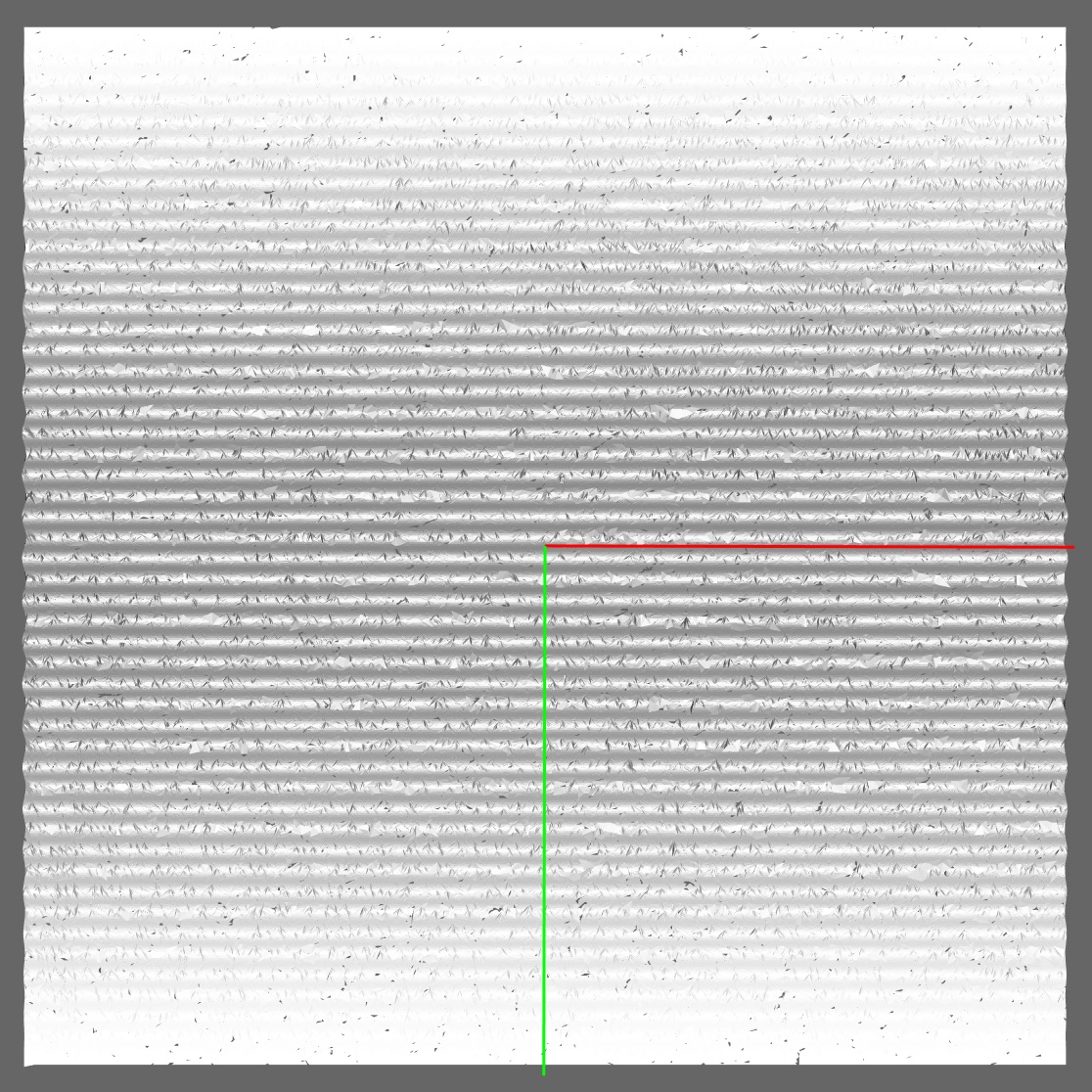}}
  \subfigure[]
  {\includegraphics[width=0.32\textwidth]{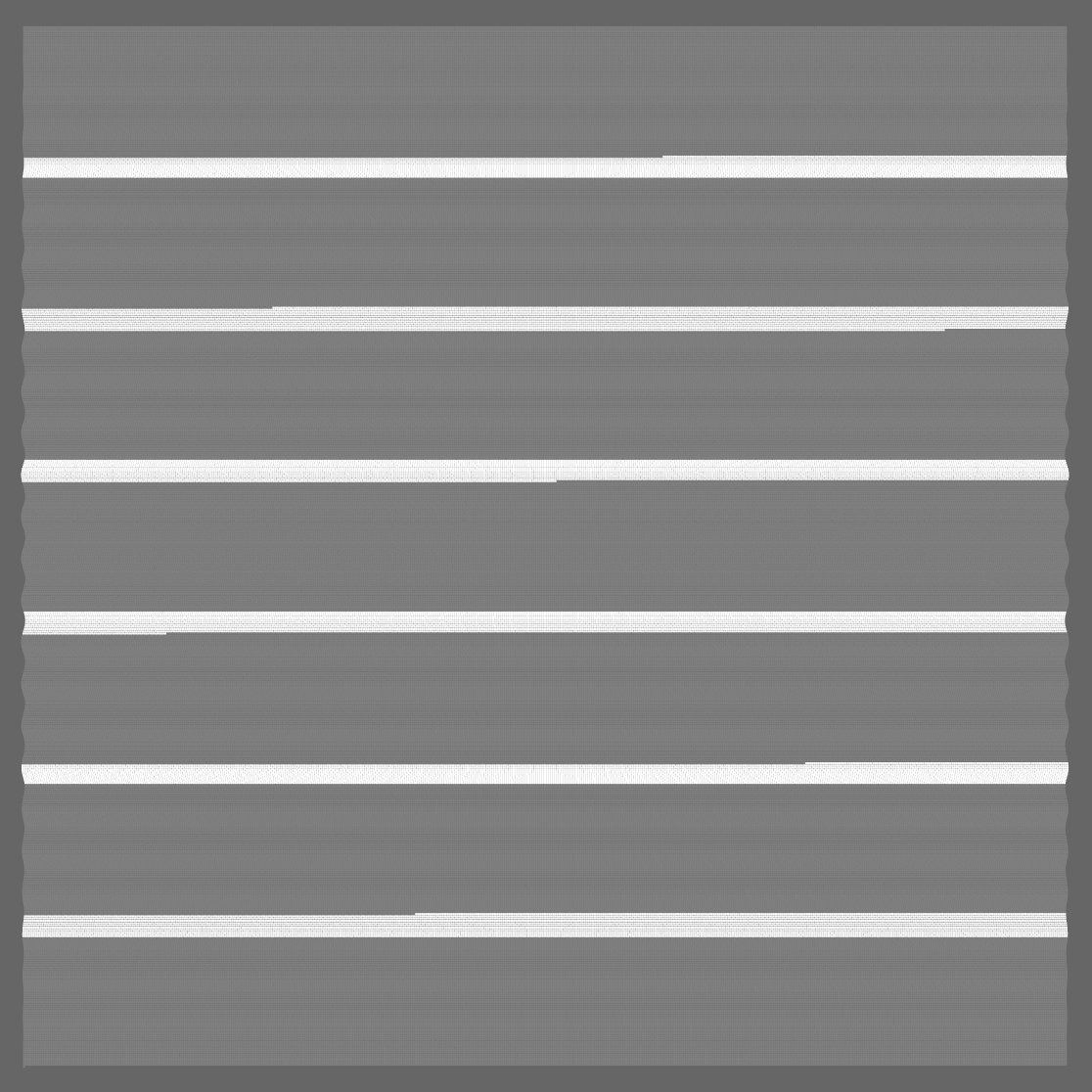}}
  \caption{The validation of the occurrence of quasi striations using idealized fracture surface with striations. (a) The constructed fracture surface with idealized striations. (b) Model $F(\phi)$ fitted to the polar diagram of $\theta_{max}^*/C$. (c) The point cloud of the constructed fracture surface with sampling distance of 1 mm. (d) The reconstructed fracture surface. The red line points in the shear direction of $0^\circ$ and the green line points in the shear direction of $90^\circ$. (e) The estimated occurrence of quasi striations shown as white stripes on the fracture surface point cloud.}
  \label{fig:striations_validation}
\end{figure}

We constructed a fracture surface with idealized striations as shown in \prettyref{fig:striations_validation}a to test the proposed method. The size of the fracture surface is $0.5\ m \times 0.5\ m$, and the 10 mm wide striations are parallel with the y axis. A point cloud of the constructed fracture surface is sampled with a sampling distance of 1 mm. The proposed method was applied on this point cloud (see \prettyref{fig:striations_validation} for details) and the estimated spatial occurrence of quasi striations, which is shown as white stripes on the fracture surface (\prettyref{fig:striations_validation}e), agrees very well with that of the constructed striations.

\begin{figure}[H]
  \centering
  \subfigure[]
  {\includegraphics[width=0.45\textwidth]{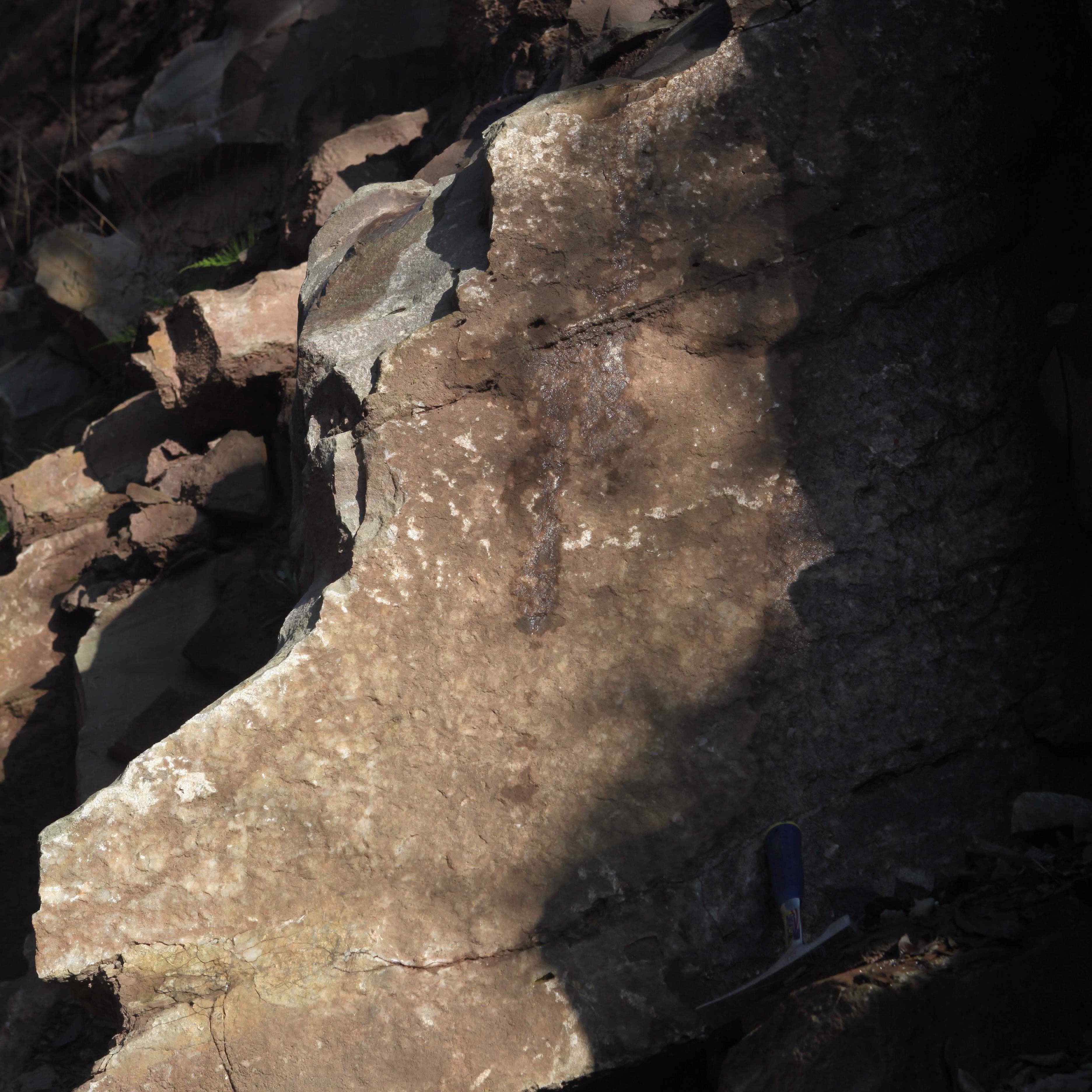}}
  \subfigure[]
  {\includegraphics[width=0.45\textwidth]{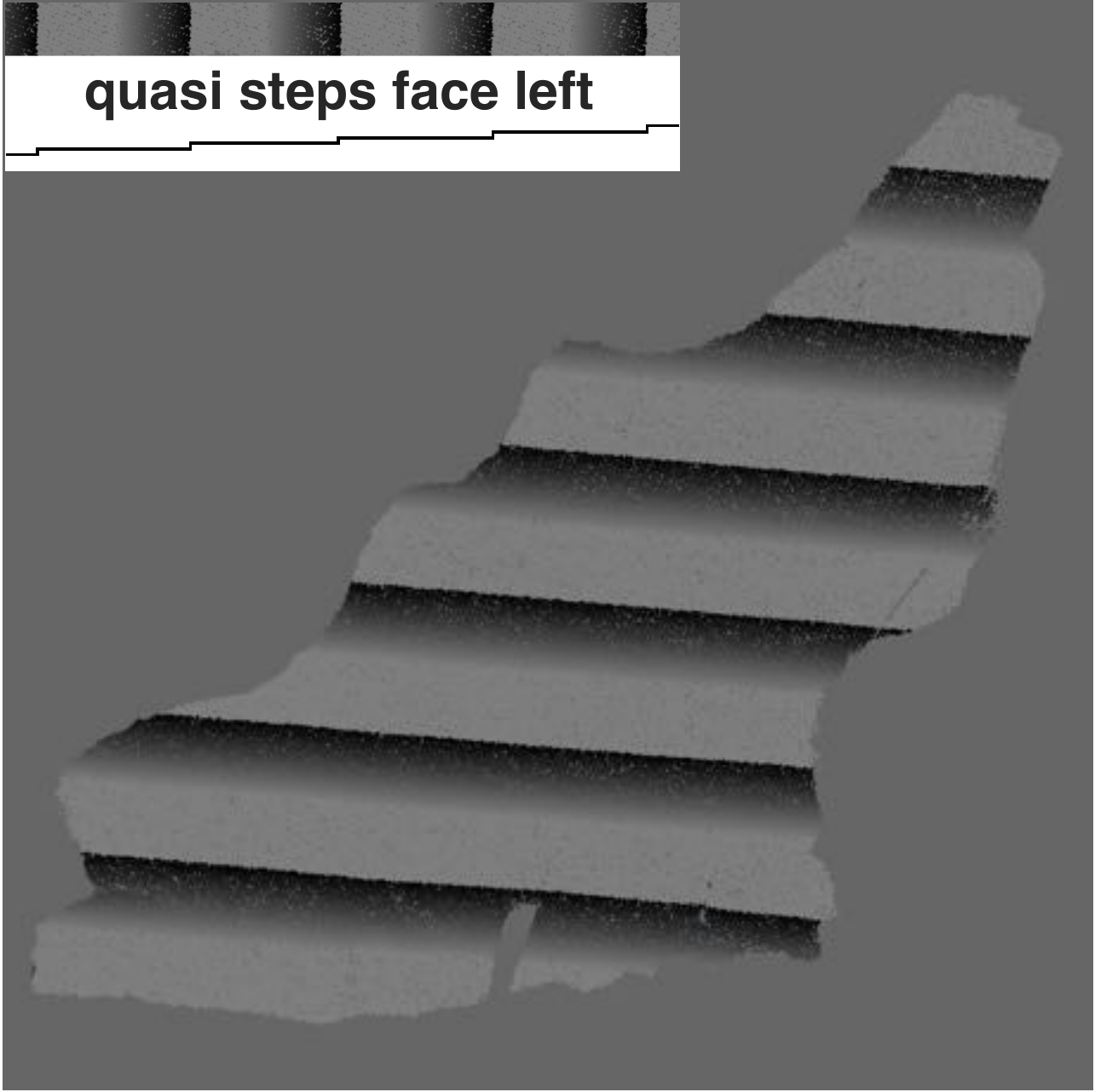}}
  \subfigure[Fault steps]
  {\includegraphics[width=0.45\textwidth]{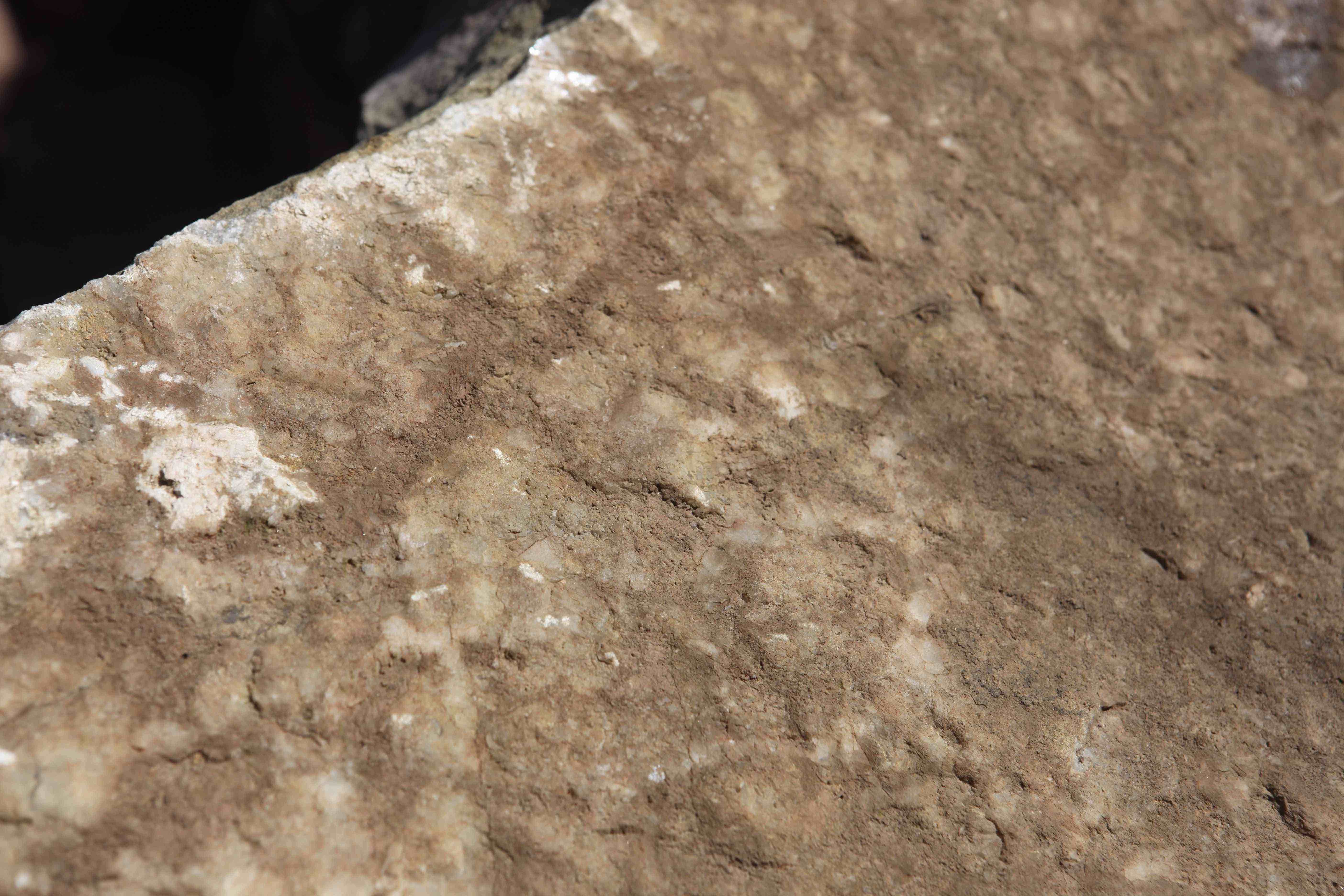}}
  \subfigure[Fault steps]
  {\includegraphics[width=0.45\textwidth]{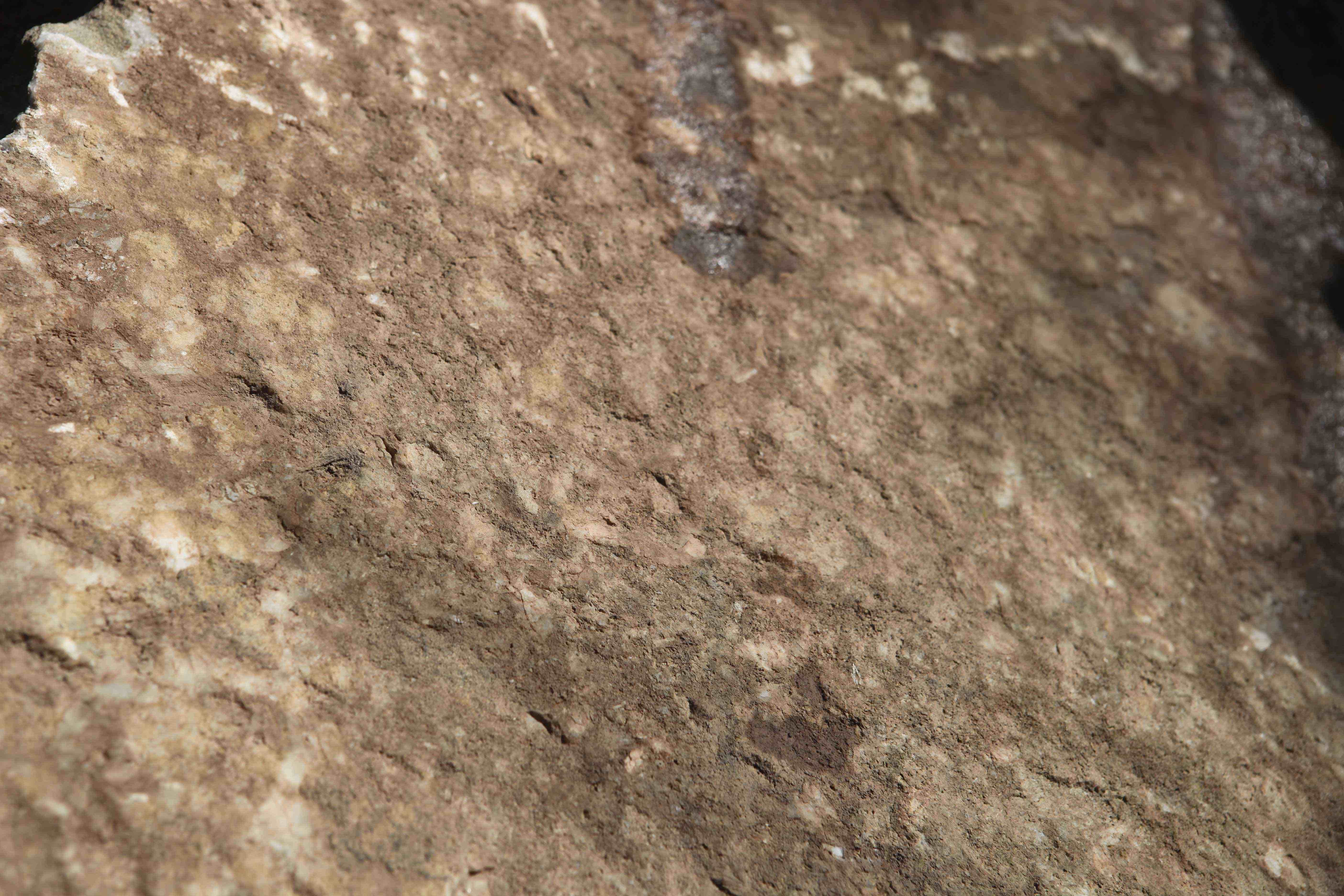}}
  \caption{The validation of the occurrence of quasi steps. (a) The picture of the fracture surface with obvious fault steps. (b) The estimated occurrence of quasi steps. (c) and (d) Pictures showing the occurrence of fault steps.}
  \label{fig:steps_validation}
\end{figure}

The proposed method was also tested on an outcrop fracture surface with obvious fault steps (\prettyref{fig:steps_validation}a). The estimated quasi steps is shown as black stripes on the fracture surface (\prettyref{fig:steps_validation}b), and the quasi steps face the direction in which the stripes' color change gradually from black to other colors (in this case, grey). Again, the estimated spatial occurrence of quasi steps agrees well with the real occurrence of fault steps shown in \prettyref{fig:steps_validation}c and \prettyref{fig:steps_validation}d. Those two examples clearly demonstrate the effectiveness of the proposed method.

\begin{figure}[H]
  \centering
  \subfigure[]
  {\includegraphics[width=0.99\textwidth]{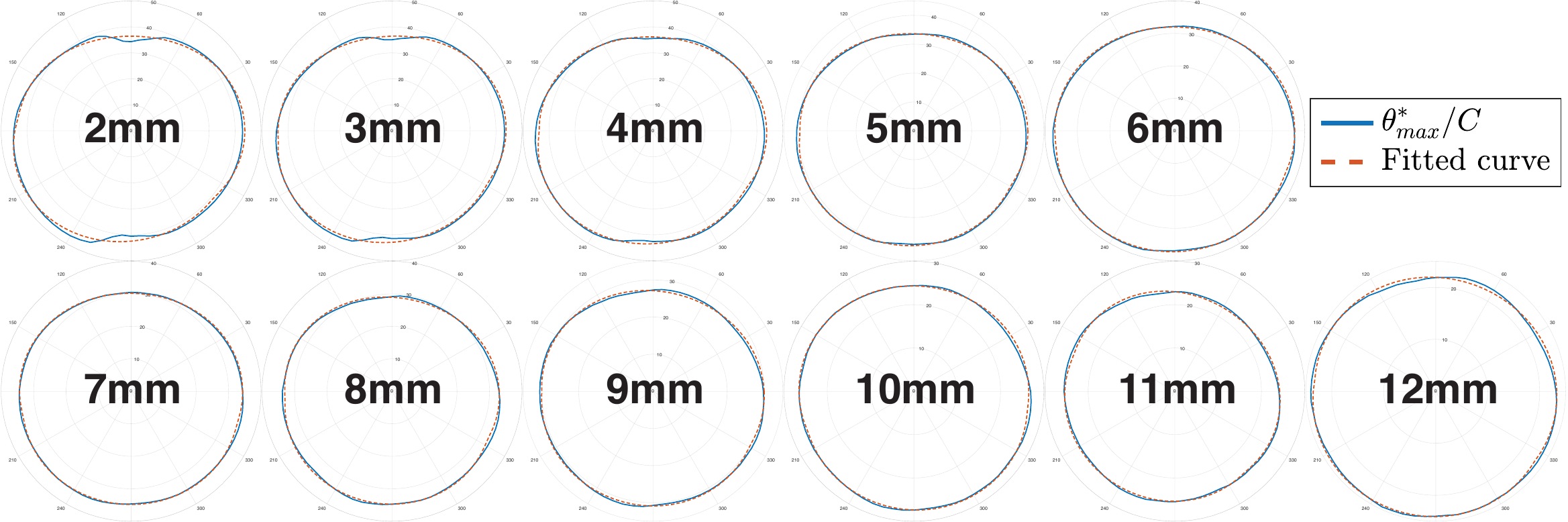}}
  \subfigure[]
  {\includegraphics[width=0.99\textwidth]{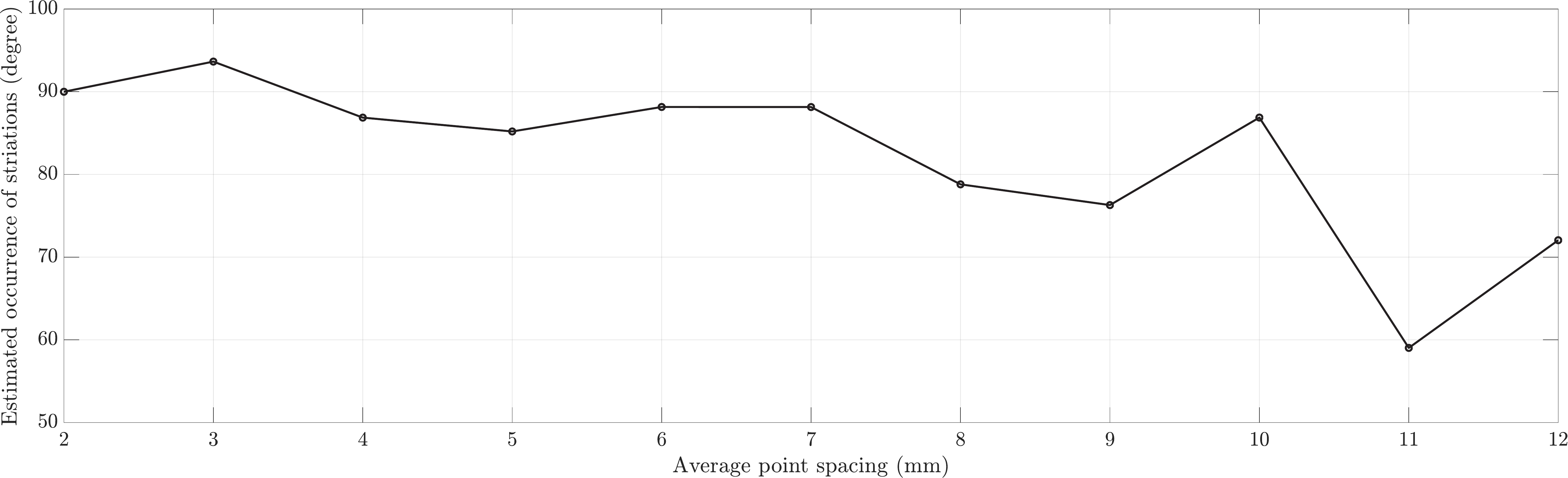}}
  \subfigure[]
  {\includegraphics[width=0.99\textwidth]{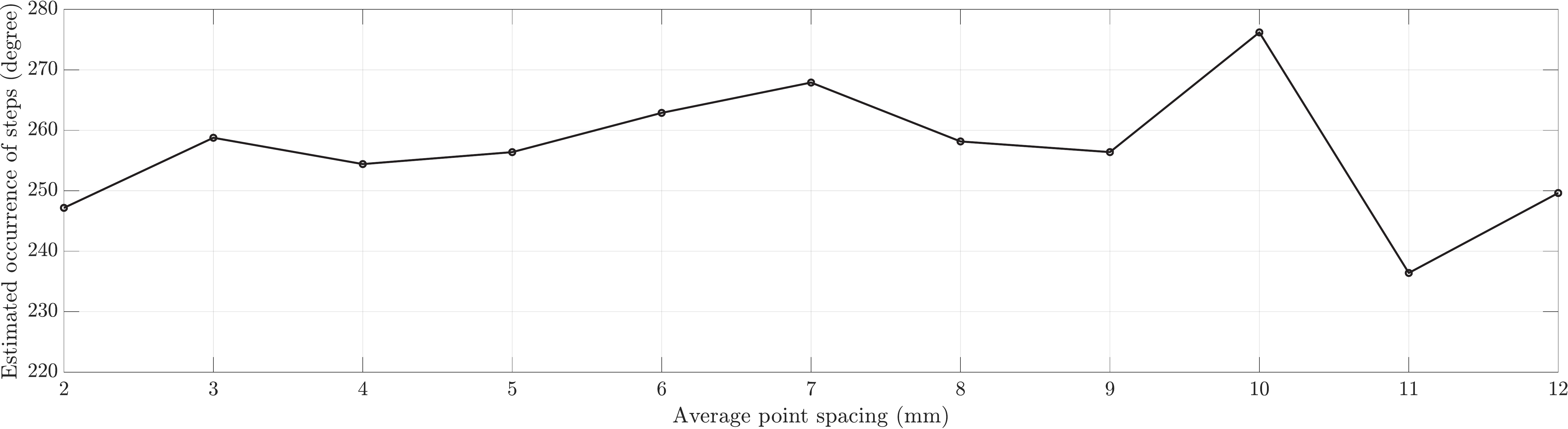}}
  \caption{Method performance analysis for the fracture surface in \prettyref{fig:before_n_after_triangulation}a with average point spacing of the point cloud ranging from 2 mm to 12 mm. (a) Models fitted to the polar diagrams of $\theta_{max}^*/C$. (b) Estimated occurrence (in the reference system shown in \prettyref{fig:after_triangulation_n_polar_diagram}a) of striations ($\phi_f$) and (c) steps ($\phi_g$) with different average point spacing of the point cloud.}
  \label{fig:results_vs_point_density}
\end{figure}

To analyze the performance of the proposed method with different point densities of the point cloud, we applied this method on the same fracture surface (\prettyref{fig:before_n_after_triangulation}a and \prettyref{fig:steps_validation}a) with average point spacing ranging from 2 mm to 12 mm. \prettyref{fig:results_vs_point_density}a shows that the anisotropic distribution of values $\theta_{max}^*/C$ caused by the indicating structures becomes increasingly obscure as the average point spacing increases and the point cloud is failing to capture the details of the indicating structures. Hence the increase of average point spacing will make the estimation of the occurrences of quasi striations and quasi steps unstable (\prettyref{fig:results_vs_point_density}b and c) for the proposed method which relies on the anisotropic distribution of values $\theta_{max}^*/C$. So a suitable average point spacing for the point cloud should be the size of small parts of most of the indicating structures on the fracture surface. \prettyref{fig:results_vs_point_density}b and c show that, for this fracture surface, the estimation of the occurrences of quasi striations and quasi steps are stable for average point spacing $<$ 10 mm. This means that the proposed method is stable as long as most of the indicating structures are well presented in the point cloud.

\section{Applications and discussion}
\label{sec:applications_and_discussion}

\begin{figure}[H]
  \centering
  \subfigure[]
  {\includegraphics[width=0.74\textwidth]{outcrop_point_cloud.pdf}}
  \subfigure[Quasi striations]
  {\includegraphics[width=0.74\textwidth]{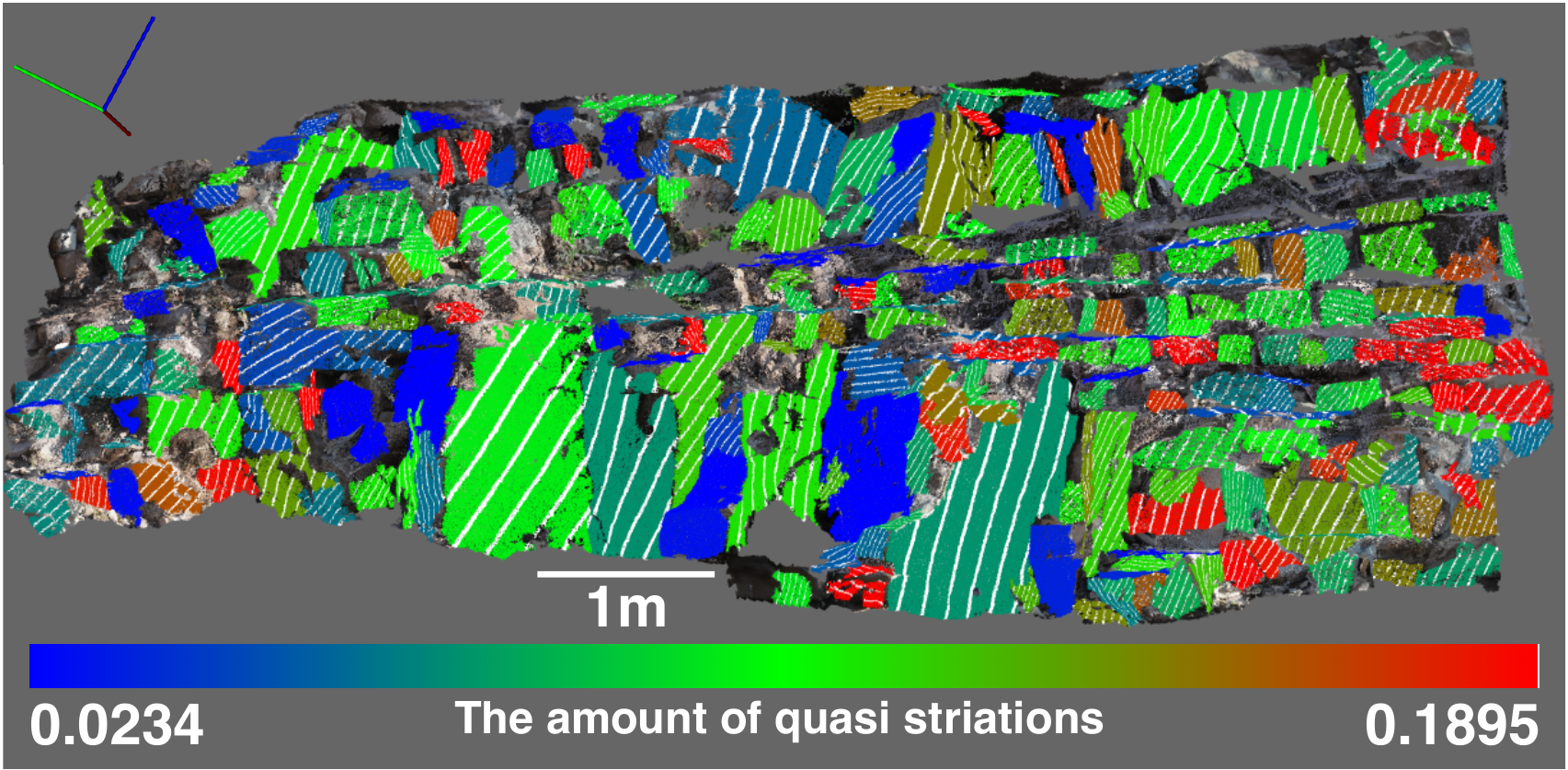}}
  \subfigure[Quasi steps]
  {\includegraphics[width=0.74\textwidth]{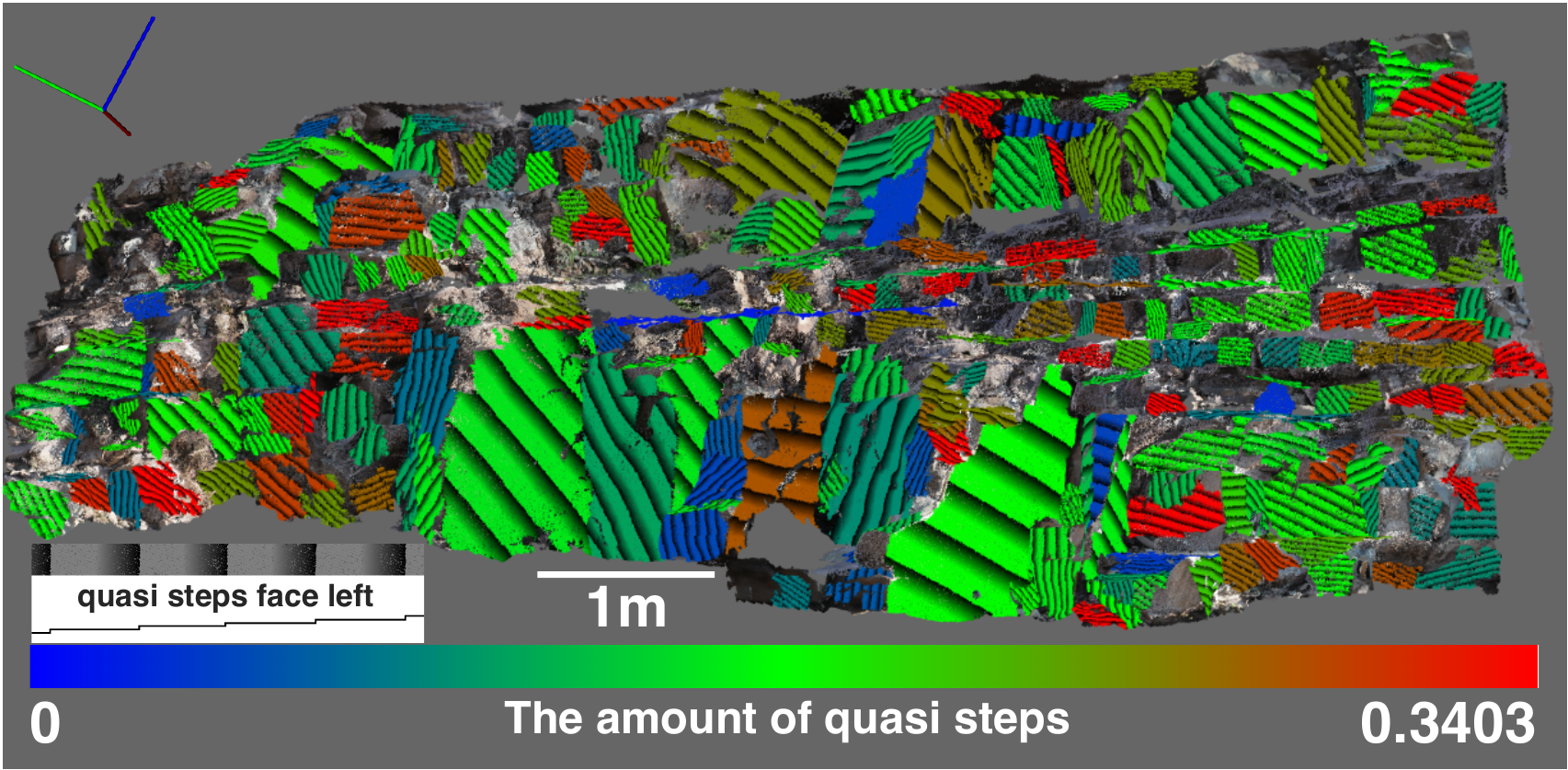}}
  \caption{(a) The outcrop point cloud. (b) The visualization of quasi striations. (c) The visualization of quasi steps. Fracture surfaces are colored based on the amount of quasi striations and quasi steps. The reference coordinate system is in the upper-left corner: the green axis points north, the red axis points east and the blue axis points vertically up.}
  \label{fig:application_striations_steps}
\end{figure}

The quasi striations and quasi steps data that indicate historical shear deformations of rock fracture has great prospect of applications, and may play a very important role in a lot of research scenarios, such as fracture network development, faulting, folding, rock mass characterization, reservoir characterization, and so on. Here we first look at the application on an outcrop located along a country road in Nanbaoxiang, Chengdu, Sichuan Province, China ($30^\circ\ 26'\ 42.09''$ N, $103^\circ\ 12'\ 17.73''$ E).  The outcrop mainly comprises thin to thick layered sandstone and the point cloud of this outcrop is shown in \prettyref{fig:application_striations_steps}a. Fracture surfaces on the outcrop are colored based on the amount of quasi striations and quasi steps. The occurrence of quasi striations is shown as white stripes on the fracture surface (\prettyref{fig:application_striations_steps}b) and the occurrence of quasi steps is shown as black stripes on the fracture surface (\prettyref{fig:application_striations_steps}c). The quasi steps face the direction in which the stripes' color change gradually from black to other colors that indicate the amount of quasi steps as stated above.

The visualization of quasi striations and quasi steps data on the outcrop gives an intuitive idea of how the rock mass was deformed. For example, fractures with less shear deformation indicators (quasi striations and quasi steps) seem to be mostly facing left as shown in \prettyref{fig:application_striations_steps}b --- in other words, the rock blocks may have undergone tensile stress in this direction during the formation of those fractures, while fractures facing right may have undergone shear stress. From the occurrence of quasi striations and quasi steps we know that fractures with similar occurrences have similar shear deformation directions (though the amount of shear deformation indicators may vary), which is consistent with the relatively stable geological background stress. Combining the occurrence of quasi striations and quasi steps of fractures facing right, we know the exact direction of the shear deformation on those fractures, and we know that there may be a shear component in the stress applied on fractures facing left.

\begin{figure}[H]
  \centering
  \subfigure[Quasi striations]
  {\includegraphics[width=0.78\textwidth]{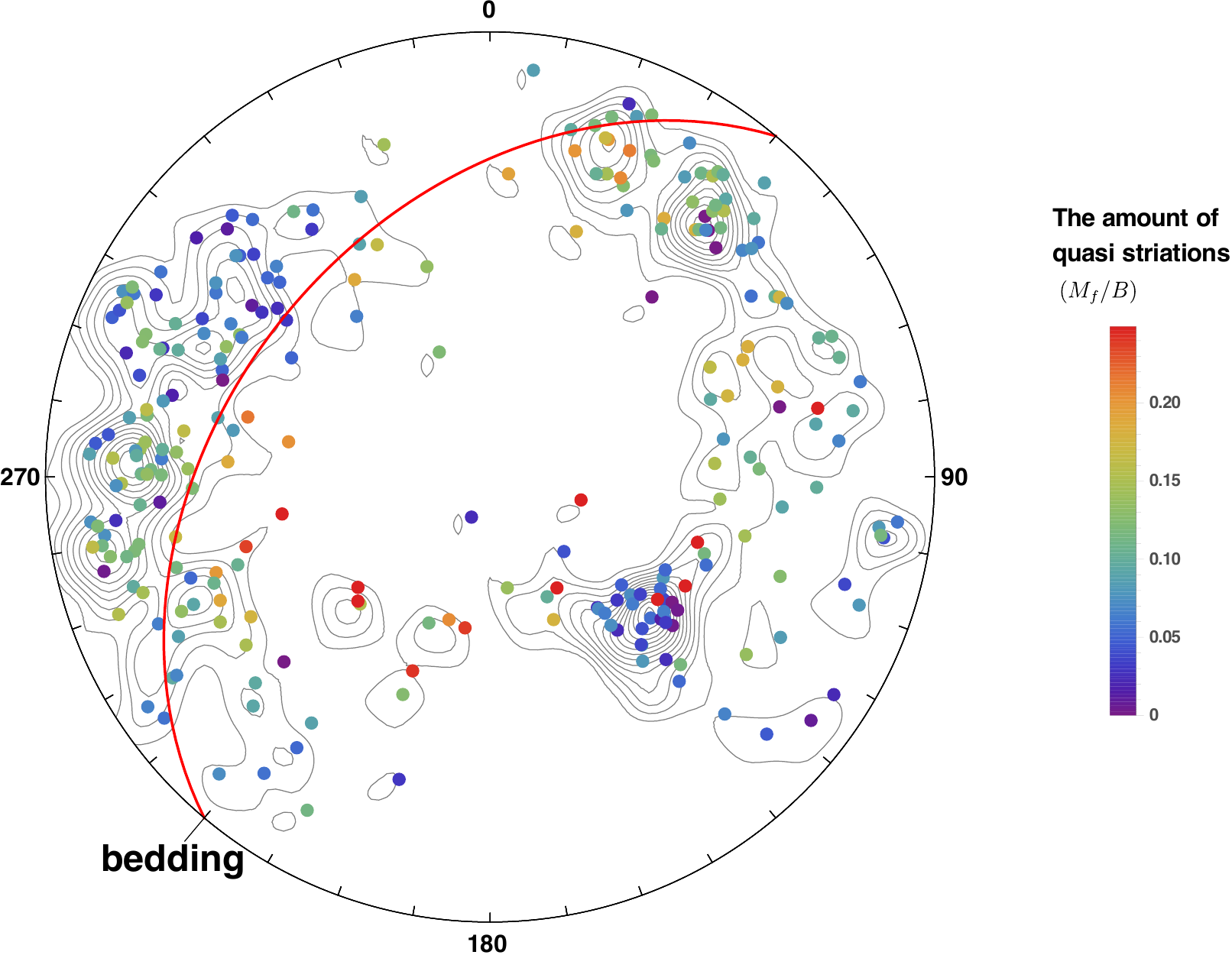}}
  \subfigure[Quasi steps]
  {\includegraphics[width=0.78\textwidth]{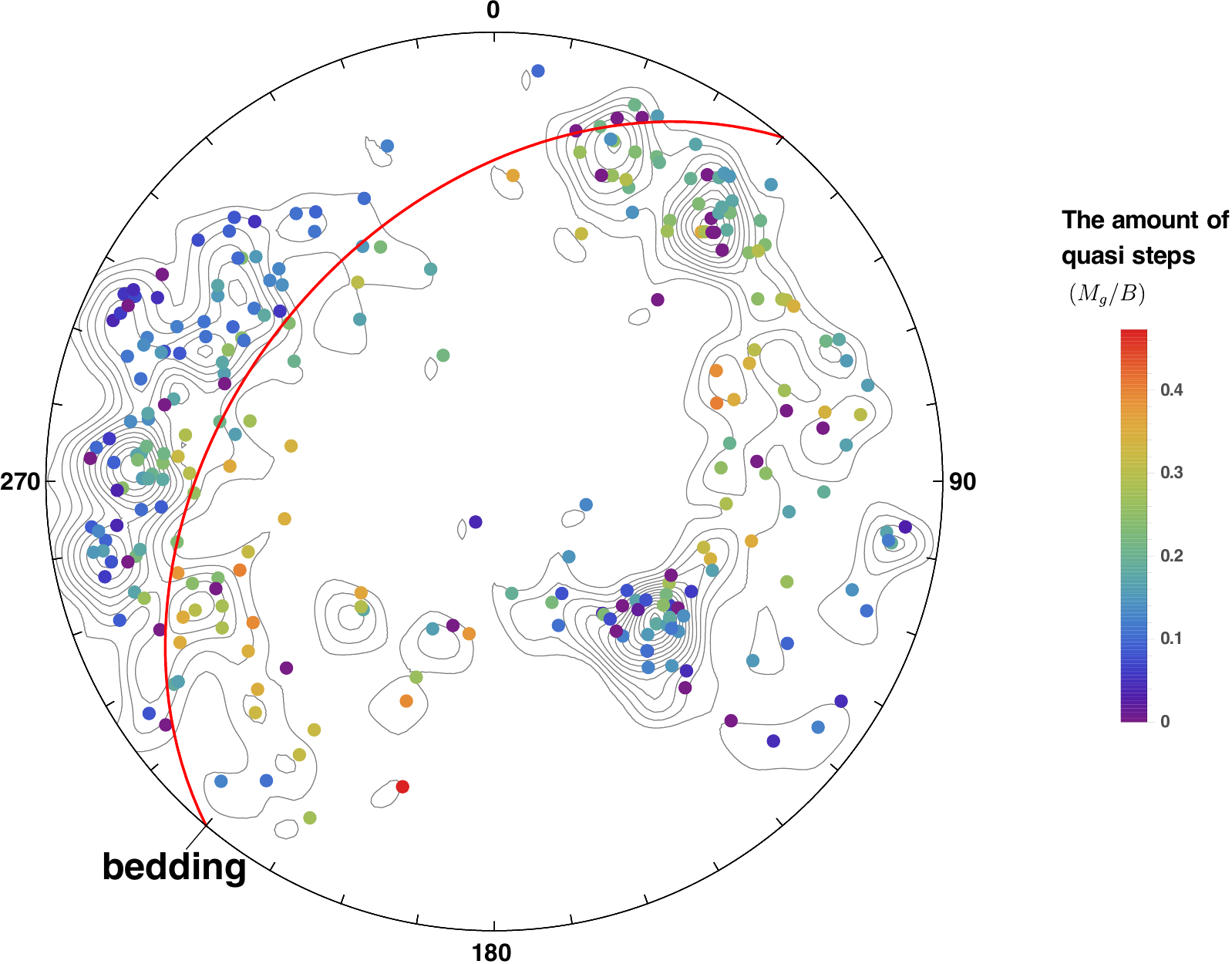}}
  \caption{Pole density plot with the amount of (a) quasi striations and (b) quasi steps.}
  \label{fig:contour_striations_steps}
\end{figure}

In addition, the amount of quasi striations and quasi steps can be integrated into the pole density plot (\prettyref{fig:contour_striations_steps}). Both \prettyref{fig:contour_striations_steps}a and b indicate that the rock fractures can be grouped into two: group one consists of fractures whose stereographic plot poles are near the arc of the bedding and group two consists of fractures whose stereographic plot poles are between the arc of the bedding and the plot border. Fractures of group one are more perpendicular to the bedding surface and are more sheared. Fractures of group two are more vertical and are less sheared. The geological history behind \prettyref{fig:contour_striations_steps}a and b may be that fractures of group one were formed as conjugated shear fractures when the bedding surfaces were still horizontal, then the strata were tilted and the geological stress was loaded on fractures of group one, which resulted in directional preference of local tensile stress (hence the preference of occurrence of tensile fractures) and the reactivation and formation of shear fractures roughly perpendicular to those tensile fractures. If the geological history is true, it demonstrates two different deformation patterns of the rock mass with and without preexisting fractures and that the distribution, occurrence and mode of new fractures are strictly controlled by preexisting fractures, as reported in the development of joints and faults in the Mount Abbot quadrangle, Sierra Nevada, California \citep{Segall1983a,Segall1983b,Martel1988,BURGMANN1994}. Thus the importance of preexisting fractures should be emphasized in modeling the development of fracture systems.

The application of quasi striations and quasi steps data will improve our understanding of a lot of fracturing-related phenomena and help us build models to predict more accurately. Further research should focus on the application of quasi striations and quasi steps data on more sophisticated problems such as faulting, folding, and so on.

\section{Conclusions}

A quantitative method to derive historical shear deformations of rock fractures from DOMs was proposed in this paper. After the extraction of individual fracture surfaces from outcrop point clouds, the fracture surfaces are reconstructed using a triangulation algorithm and the shear strength parameter $\theta_{max}^*/C$ is calculated for shear directions all around the fracture surface. A theoretical model that combines the effects of isotropic ``base'' shear strength, quasi striations and quasi steps on the shear strength parameter of rock fracture is fitted to the calculated $\theta_{max}^*/C$, thus the amount and occurrence of quasi striations and quasi steps are estimated, and the historical shear deformations can be inferred.

The validity of the proposed method was proved by testing it on a constructed fracture surface with idealized striations and a fracture surface with clear fault steps. The application of this method on an example outcrop shows an intuitive idea of how the rock mass was deformed and that the distribution, occurrence and mode of new fractures are strictly controlled by preexisting fractures. The importance of preexisting fractures should be emphasized in modeling the development of fracture systems.

There is great prospect of applications of the proposed method and the quasi striations and quasi steps data. Further research should focus on the application of this method on problems such as faulting, folding, etc.

\section*{Acknowledgement}

This study was funded by the Chinese National Science and Technology Major Project (2017ZX05008-001) and National Natural Science Foundation of China (41872214), and was partially funded by the Special Fund from Zhejiang Provincial Government, China (zjcx. 2011 No. 98). Xin Wang would like to acknowledge the support of Institute for Pure and Applied Mathematics for his long-term visits. We would like to thank the editor and the anonymous reviewers for their valuable comments and suggestions, which have improved the paper. Readers can find the example outcrop point cloud data from https://figshare.com/articles/outcrop-point-cloud/6936437/1 \citep{Wang2018}.


\bibliography{elsarticle-template.bib}

\end{document}